\documentclass[pra, letterpaper, twocolumn, superscriptaddress, showpacs]{revtex4}

\newcommand{\FigDirectory}{.}

\usepackage{amsmath}                          % better math, text control.
\usepackage{amsfonts}                         % extended math fonts.
\usepackage{amssymb}                          % names amsfont symbols.
\usepackage{amsthm}                           % theorem environments

\usepackage{dsfont}                           % doublestroke \CC, etc.
\usepackage[usenames]{color}                  % font colors.

\usepackage{graphicx}                         % ps, pdf figs.
\usepackage{epstopdf}                         % eps to pdf if detected.
\usepackage{subfigure}                        % for subfigures
\usepackage[pdfpagelabels,pdftex,bookmarks,breaklinks]{hyperref}
\hypersetup{colorlinks,citecolor=blue,urlcolor=blue,linkcolor=black}
\hypersetup{pdfstartview=FitH,pdfpagemode=None}
\hypersetup{pdftitle=Adiabatic\ Topological\ Quantum\ Computing}

\usepackage{datetime}                         % formatting date, time

\input{Qcircuit}

\renewcommand{\>}{\rangle}

\begin{document}

%%%%%%%%%%%%%%%%%%%%%%%%%%%%%%%%%%%%%%%%%%%%%%%%%%%%%%%%%%%
% AJL new commands: post "begin document"

\makeatletter
\def\grabtimezone #1#2#3#4#5#6#7#8#9{\grabtimezoneB}
\def\grabtimezoneB #1#2#3#4#5#6#7{\grabtimezoneC}
\def\grabtimezoneC #1#2'#3'{#1#2#3 UTC}
\def\timezone{\expandafter\grabtimezone\pdfcreationdate}
\makeatother

%\timezone

%%%%%%%%%%%%%%%%%%%%%%%%%%%%%%%%%%%%%%%%%%%%%%%%%%%%%%%%%%%%%%%%%%%%%%%%%%%%%%%
% Title & Authors

% Physical Review Style:
%
\title{Adiabatic topological quantum computing}

\author{Chris \surname{Cesare}}
\email[]{chris.cesare@gmail.com}
\affiliation{Center for Quantum Information and Control,
             University of New Mexico,
             Albuquerque, NM, 87131, USA}
\affiliation{Department of Physics and Astronomy,
             University of New Mexico,
             Albuquerque, NM, 87131, USA}
\author{Andrew J. \surname{Landahl}}
\email[]{alandahl@sandia.gov}
\affiliation{Advanced Device Technologies,
             Sandia National Laboratories,
             Albuquerque, NM, 87185, USA}
\affiliation{Center for Quantum Information and Control,
             University of New Mexico,
             Albuquerque, NM, 87131, USA}
\affiliation{Department of Physics and Astronomy,
             University of New Mexico,
             Albuquerque, NM, 87131, USA}
\author{Dave \surname{Bacon}}
\email[]{dabacon@gmail.com}
\altaffiliation[current affiliation:~]{Google, Inc.,
             651 N. 34th St.
             Seattle, WA, 98103, USA}
\affiliation{Department of Computer Science and Engineering,
              University of Washington,
              Seattle, WA, 98195, USA}
\affiliation{Department of Physics,
              University of Washington,
              Seattle, WA 98195, USA}
\author{Steven T. \surname{Flammia}}
\email[]{sflammia@physics.usyd.edu.au}
\affiliation{Centre for Engineered Quantum Systems,
             School of Physics,
             The University of Sydney,
             Sydney, NSW 2006, Australia}
\author{Alice \surname{Neels}}
\email[]{aliceen@gmail.com}
\altaffiliation[current affiliation:~]{Splunk, Inc.,
             250 Brannan St., 1st Floor,
             San Francisco, CA 94107, USA}
\affiliation{Department of Computer Science and Engineering,
              University of Washington,
              Seattle, WA, 98195, USA}

%\date[\bf DRAFT:\rm\ ]{\today, \currenttime\ \timezone}

%%%%%%%%%%%%%%%%%%%%%%%%%%%%%%%%%%%%%%%%%%%%%%%%%%%%%%%%%%%%%%%%%%%%%%%%%%%%%%%
% Abstract

\begin{abstract}

Topological quantum computing promises error-resistant quantum computation
without active error correction.  However, there is a worry that during the
process of executing quantum gates by braiding anyons around each other,
extra anyonic excitations will be created that will disorder the encoded
quantum information.  Here we explore this question in detail by studying
adiabatic code deformations on Hamiltonians based on topological codes,
notably Kitaev's surface codes and the more recently discovered color codes.
We develop protocols that enable universal quantum computing by adiabatic
evolution in a way that keeps the energy gap of the system constant with
respect to the computation size and introduces only simple local Hamiltonian
interactions.  This allows one to perform holonomic quantum computing with
these topological quantum computing systems.  The tools we develop allow one
to go beyond numerical simulations and understand these processes
analytically.   

\end{abstract}

% For Physical Review Only:
%
\maketitle

%%%%%%%%%%%%%%%%%%%%%%%%%%%%%%%%%%%%%%%%%%%%%%%%%%%%%%%%%%%%%%%%%%%%%%%%%%%%%%%
% Body

%%%%%%%%%%%%%%%%%%%%%%%%%%%%%%%%%%%%%%%%%%%%%%%%%%%%%%%%%%%%%%%%%%%%%%%%%%%%%%%
% Section
%
\section{Introduction}
\label{sec:introduction}

There are many approaches to constructing a quantum computer.  In addition
to the numerous different physical substrates available, there are a
plethora of different underlying computational architectures from which to
choose.  Two major classes of architectures can be distinguished: those
requiring a substantial external active control system to suppress
errors~\cite{Shor:1995a, Steane:1996a, Preskill:1998a}, and those whose
underlying physical construction eliminates much, if not all, of the need
for such a control system~\cite{Kitaev:1997a, Farhi:2000a}.  The first class
of architectures strives to minimize the control resources needed to quantum
compute fault-tolerantly.  The second class of architectures strives to
minimize the complexity of systems that enable fault-tolerant quantum
computation intrinsically. Here we focus on the latter class of
architectures and address the question: ``How does one quantum compute on a
system protected from decoherence by a static (\textit{i.e.},
time-independent) Hamiltonian?'' We present a solution that adiabatically
interpolates between static Hamiltonians, each of which protects the quantum
information stored in its ground space.  Since each of these ground spaces
can be described as a quantum error-correcting codespace, we call this
process \emph{adiabatic code deformation} \cite{Oreshkov:2009a,
Bacon:2009a}.  This procedure amounts to a simulation of the
measurement-based process of code deformation employed in the first class of
architectures~\cite{Dennis:2002a, Raussendorf:2006a, Bombin:2007e,
Fowler:2009, Bombin:2011a, Horsman:2012, Bonderson:2013a}.  We further
show that this procedure preserves the energy gap of the system throughout
the evolution.

While previous work has made reference to adiabatic evolutions as a method
for performing topological quantum computation~\cite{Nayak:2008a}, our work
can be seen as making the assumptions of adiabatic evolution explicit for
certain models of topological quantum computers.  In contrast, for example,
to topological quantum computing in fractional quantum Hall systems where
even the ground state of the system is subject to debate, our models are
exactly solvable and simple.  Similar work has been performed for Kitaev's
honeycomb model by Lahtinen and Pachos~\cite{Lahtinen:2009a}, who examined
the adiabatic transport of vortices in Kitaev's honeycomb lattice model
numerically.  Here, we are able to investigate these issues analytically.

Our results marry three different lines of research, which we now describe.
The first is the idea originated by Kitaev~\cite{Kitaev:1997a} that quantum
information can be protected from decoherence by encoding it into the
degenerate ground space of a many-body quantum system.  In particular,
Kitaev suggested a family of systems such that each system has a ground
space equivalent to a quantum error-correcting codespace.  Moreover, each of
these ground spaces is separated from its first-excited space by an energy
gap---a gap that does not shrink with the system size (\textit{i.e.}, the
gap is ``constant'').

In Kitaev's original construction, the quantum error-correcting code also
possesses a topological property that makes the distance of the code grow
with the number of qubits in the system.  This implies that any local
perturbing interaction will only split the energy of a degenerate ground
state by an exponentially small amount in the size of the
system~\cite{Bravyi:2010a}.  Information encoded into the ground space
should therefore remain well-protected from the detrimental effects of
decoherence.  Further, if one immerses the system in a bath with a
temperature lower than that of the energy gap in the system, then one should
expect a suppression of thermal excitations out of the ground space.  The
decay rate of the quantum information encoded into the ground space is {\em
not} set by a length scale in the system, but instead the lifetime scales as
$\exp(c \beta \Delta)$ where $\beta$ is the inverse temperature,  $\Delta$
is the energy gap of the Hamiltonian, and $c$ is a
constant~\cite{Alicki:2009a}.  Crucially, this implies that the lifetime of
the information is exponentially lengthened as a function of the inverse
temperature.  While one does not obtain, using Kitaev's original idea, a
method for protecting quantum information with a lifetime that grows with
the size of the system---a hallmark of ``self-correcting'' quantum memories
\cite{Dennis:2002a, Bacon:2006a}---for a suitably low temperature, the
information lifetime will be long enough for all practical purposes. Thus,
via the use of a static many-body Hamiltonian, Kitaev proposed that quantum
information could be protected without resorting to active quantum
error-correcting algorithms.

Following Kitaev's introduction of this idea, numerous authors put forward
similar approaches.  Many of these ideas stayed within the realm of
topological protection~\cite{Freedman:2003b, Freedman:2003c, Brink:2003a,
Kitaev:2006b, Freedman:2006a, Nayak:2008a}, but others explored energetic
protection without reference to topological ideas~\cite{Barnes:2000a,
Bacon:2001b, Weinstein:2005a, Bacon:2008a}.  Here we will focus on the
topological models, but many of our results apply in the more general
setting.

Kitaev noted in his original proposal that the excited states of his
Hamiltonian act as particles with exotic statistics.  In particular, he
showed that the excitations were quasiparticles called
\emph{anyons}~\cite{Wilczek:1982a}---particles that exist in two spatial
dimensions that exhibit statistics different from fermions and bosons and
which interact by braiding around one another in spacetime---an interaction
that only depends on the topology of the anyon worldlines.  These
excitations not only describe errors in the codespace but can also be
thought of as quantum information carriers in their own right.  Indeed for
some many-body Hamiltonians, it is possible to have \emph{nonabelian} anyons
(anyons whose braidings do not commute) that perform universal quantum
computation in the label space of the anyons.  This is known as
\emph{topological quantum computing}~\cite{Kitaev:1997a, Freedman:2002a,
Freedman:2002b, Koenig:2010b}, the principal model of quantum computing we
will consider here.

In a topological quantum computation, one creates anyons from the vacuum,
braids them around one another in spacetime, fuses them together, then
records their label types.  Although the topological nature of the anyonic
interaction provides a degree of control robustness, it is not immediately
clear why the processes of anyon creation and fusion could not create new
unwanted anyons.  Such anyons could in turn wander and disrupt the desired
braid.  The initialization process in particular is quite
subtle~\cite{Koenig:2010a}.  Moreover, there will likely be a background
of thermal anyons and anyons arising from material defects which could also
disorder the quantum computation.  On top of all of this, even if a
spacetime braid is topologically correct, the mere act of moving anyons
around at any nonzero speed has the potential to generate new excitations
because the adiabatic approximation is not exact.  Measurement-based
topological quantum computation~\cite{Bonderson:2008a, Bonderson:2008b} has
the potential to overcome this last problem, but the other problems remain.
In summary, the great merit of topological quantum computation is that the
``only'' thing that can corrupt it is uncontrolled anyons---the problem is
that there are many ways that uncontrolled anyons can arise.  Even something
as seemingly innocuous as a lack of complete knowledge of the system's
Hamiltonian could do this because it could lead to anyons being trapped or
leaking out of the system unbeknownst to the computer operator
\cite{Nayak:2008a}.  We do not claim to address every possible adversarial
scenario for topological quantum computation here; our focus is on
constructing an architecture that limits the chances for uncontrolled
anyons to appear.

The second line of research relevant to our proposal is the recent use of
code deformations to perform quantum computation on topological quantum
error-correcting codes~\cite{Dennis:2002a, Raussendorf:2006a, Bombin:2007e, 
Bombin:2011a, Koenig:2010b}.  In this approach, one works directly with the 
quantum-error correcting code used in topological quantum computing without 
introducing a Hamiltonian to provide energetic protection of the quantum 
information.  Instead, one focuses on active error correction, but performed with 
the topological quantum codes.  Consideration of such codes for quantum error
correction was first examined in detail by Dennis {\em et
al.}~\cite{Dennis:2002a}.   In this approach, qubits are arranged on a
two-dimensional surface with a boundary, resulting in a single encoded qubit
for each such surface.  In order to build a quantum computer with more than
one qubit, such surfaces are stacked on top of each other so that
transversal gates can be achieved between the neighboring surfaces.  Since
the original analysis, modifications~\cite{Raussendorf:2007b, Bombin:2007e}
of this architecture have been introduced that have considerable advantages
over the three-dimensional stacking of Dennis {\em et al.}  In these models,
one takes a surface code and ``punctures'' it by removing the quantum check
operators (stabilizer generators) from a region, creating a {\em
defect}~\cite{Bravyi:1998a}.  For each defect, one obtains an encoded qubit
with a code distance that is the minimum of the perimeter of the defect and
the distance from the defect to the nearest appropriate boundary (which may
lie on another defect).  One can show that, via a sequence of adaptive
measurements, one can {\em deform} the boundary of the defect, and, by using
suitable deformations, braid defects in such a way that logical operations
are performed between the logical qubits associated with the defects. 

The third line of research relevant to our proposal is the recent discovery
of methods to perform holonomic~\cite{Zanardi:1999a} and open-loop
holonomic~\cite{Kult:2006a} universal quantum computation in a stabilizer
code setting~\cite{Oreshkov:2009a, Bacon:2009a, Oreshkov:2009b}.  In
holonomic quantum computing, adiabatic changes of a Hamiltonian with
degenerate energy levels around a loop in parameter space induce unitary
gates on each energy eigenspace.  The enacted gate depends on geometric
properties of the Hamiltonian path and not on the exact timing used to
traverse it (to within the limits of the adiabatic approximation), thus
offering a method to avoid some timing errors.  Universal quantum
computation using holonomic methods was originally studied in
Ref.~\cite{Zanardi:1999a}.  Recently, Oreshkov {\em et al.} demonstrated a
novel manner for achieving universality within the context of fault-tolerant
quantum computing~\cite{Oreshkov:2009a}.  In particular, this result showed
how to perform gates on information encoded into a quantum stabilizer code.
Building along these lines, two of the present authors (DB and STF) have
shown how to achieve similar constructions within the context of open-loop
holonomic quantum computation~\cite{Bacon:2009a, Bacon:2010b}.  In this
setting, instead of using cyclic evolutions, one can quantum compute using
non-cyclic evolutions.  A consequence of this is a scheme known as
\emph{adiabatic gate teleportation} where one mimics gate teleportation via
a very simple interpolation between two-qubit
interactions~\cite{Bacon:2009a}.  Another consequence is that it is possible
to perform measurement-based quantum computing~\cite{Raussendorf:2001a}
using only adiabatic deformations of a Hamiltonian~\cite{Bacon:2010b}.  Further,
and more suggestive for this work, it is possible to perform holonomic quantum
computation on symmetry protected spin chains~\cite{Renes:2013a}.
Holonomic quantum computation, whether performed cyclically or
non-cyclically, should be distinguished from (universal) adiabatic quantum
computation, in which the ground state is always nondegenerate throughout
the non-cyclic adiabatic evolution~\cite{Farhi:2001a, Aharonov:2004a,
Oliveira:2005a, Mizel:2006a, Kempe:2006a}.

In this work, we combine many of the above insights into a new method for
computing on information encoded into the energy levels of a Hamiltonian.
We consider a situation where, as in the first line of research, quantum
information is encoded into the ground state of a topologically ordered
many-body system.  Rather than storing information in the label space of
anyons themselves, we consider information stored in defects, which act
somewhat like anyons, as in the second line of research.  We then
examine explicit adiabatic interpolations between Hamiltonians that simulate
code deformation, as in the third line of research.  This is all done while
keeping the energy gap in the system constant, a necessary requirement to
use these techniques to maintain the topological protection offered by these
systems.  Further, we demonstrate how to prepare quantum information into
fiducial states using adiabatic evolutions.  Some of these state-preparation
procedures are robust to error, but some (\textit{e.g.}, the preparation of
certain ``magic states''~\cite{Bravyi:2005a}) are not robust and thus
require distillation protocols.  Finally, we discuss how one can use code
deformations to facilitate measurements of certain logical operators.  We
discuss all of these procedures first within the context of Kitaev's surface
codes with defects, and then we discuss how these results can be extended to
the topological color codes \cite{Bombin:2006b}.

The systems and protocols we use are not strictly fault-tolerant. Without
active error correction, the lifetime of the codes studied are a constant
independent of the system size \cite{Alicki:2009a}. As mentioned above, here
we rely on a coupling to a cold (with respect to the gap) thermal bath,
which suppresses the creation of errors exponentially in the size of the
gap. We retain robustness to things like control errors by virtue of the
holonomic nature of the logical operations we implement, and robustness to
correlated fluctuations induced by the environment by keeping defects
well-separated during braiding. Once the environment creates an excitation,
it is free to wander and corrupt the computation. We prevent the environment
from doing this by ensuring that it is cold, and we prevent ourselves from
introducing excitations accidentally by carefully designing our procedures.

%%%%%%%%%%%%%%%%%%%%%%%%%%%%%%%%%%%%%%%%%%%%%%%%%%%%%%%%%%%%%%%%%%%%%%%%%%%%%%%
% Section
%
\section{Surface codes with defects}

We begin by working with a simple class of surface codes with defects to
establish the main ideas behind our procedures.  In Section~\ref{sec:color}
we extend these ideas to the topological color codes.  We assume that the
reader is familiar with the theory of stabilizer
codes~\cite{Gottesman:1997a}, toric codes \cite{Kitaev:1997a}, and surface
codes \cite{Bravyi:1998a}, which are specializations of toric codes to
bounded planar surfaces. However, we review these subjects to set our
notation.  

\begin{figure}[h]
\begin{center}
\includegraphics[width=1.0\columnwidth]{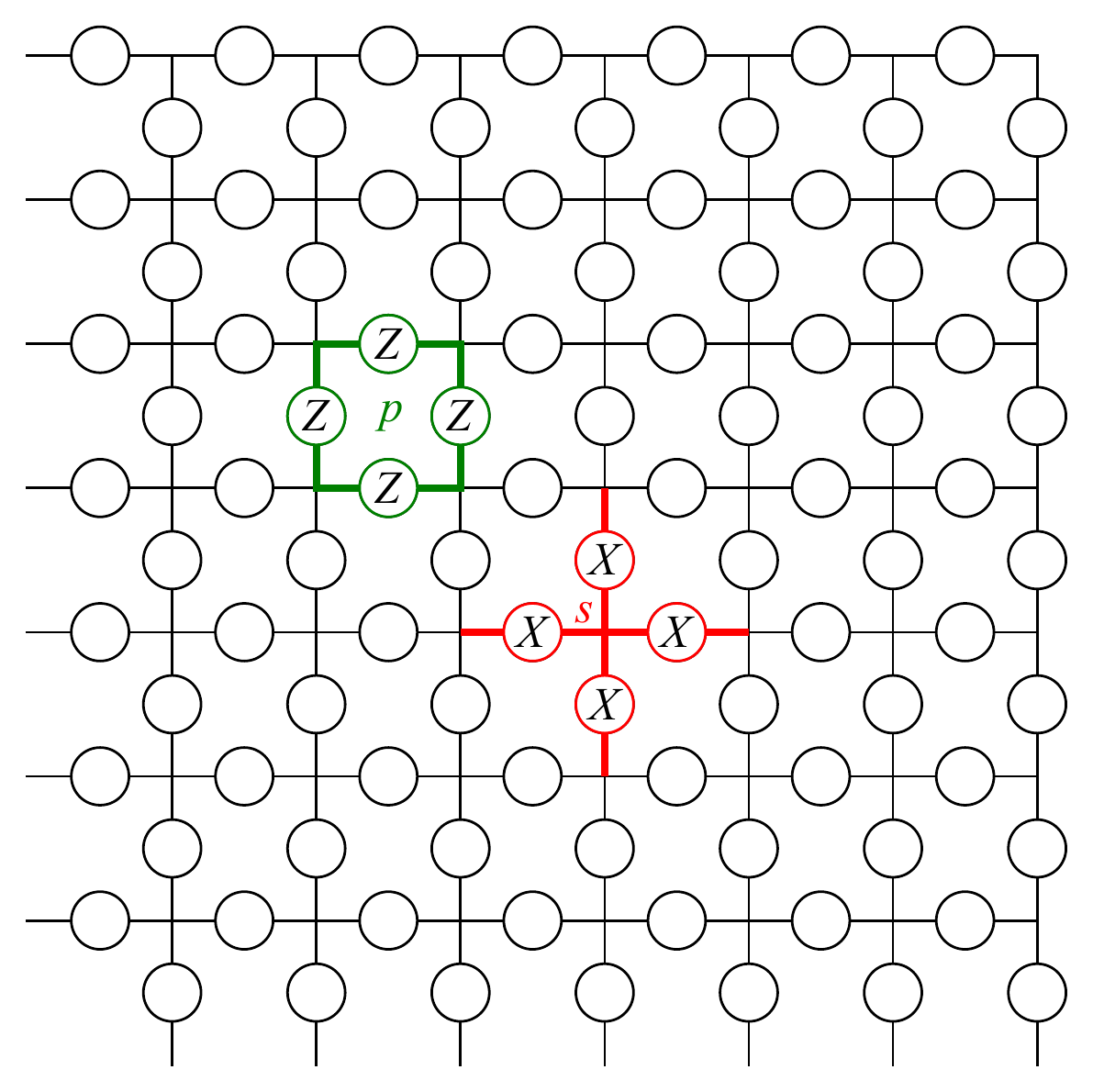}
\caption{Stabilizer generators (checks) for the surface code.  An example of
a plaquette check $S_p$ and a vertex check $S_v$.\label{fig:stabilizers}}
\end{center}
\end{figure}

\begin{figure}[t]
\begin{center}
\includegraphics[width=0.95\columnwidth]{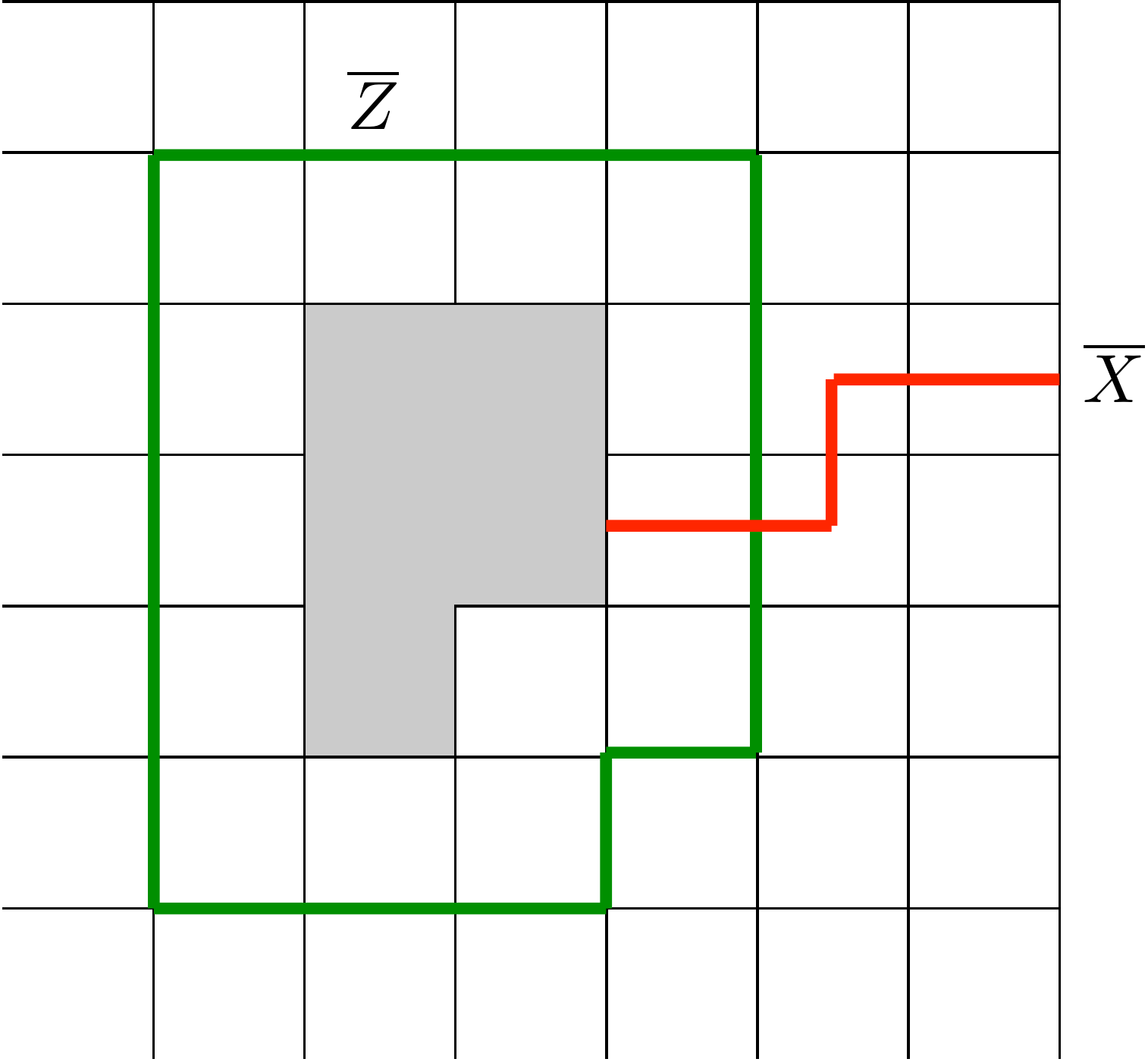}
\caption{A smooth ($Z$-type) defect.  A logical $Z$ operator is defined by a
closed loop of $Z$s on the lattice that surrounds the defect and a logical
$X$ operator is defined by a connected path of $X$s on the dual lattice from
the defect to a {\em smooth} (\emph{$Z$-type}) boundary.  Here we depict the
removed region by removing that part of the lattice; this simply indicates
that the code of the system factors into a code in the drawn region and a
code inside of the defect.}
\label{fig:smoothdefect}
\end{center}
\end{figure}

Let ${\mathcal L}$ be a two-dimensional square lattice that is $l$ edges (or
links) wide and $l$ edges tall, with the leftmost $l$ vertical edges and
bottommost $l$ horizontal edges removed. (Other lattices are possible; we
make this restriction only to be concrete.)  We call the sides of the
lattice with the edges removed the \emph{rough} or \emph{$X$-type}
boundaries and the other sides the \emph{smooth} or \emph{$Z$-type}
boundaries; see Fig.~\ref{fig:smoothdefect}.  A qubit is associated with
each edge of the lattice so that there are $2l^2$ qubits in total. For each
plaquette (or face), $p$, of the lattice, define the {\em plaquette}
operator $S_p=\bigotimes_{e \in \partial p} Z_e$ where $\partial p$ denotes
the edges bounding the plaquette and $Z_e$ is the Pauli $Z$ operator acting
on the qubit at edge $e$.  In other words $S_p$ acts as the tensor product
of $Z$ operators on the qubits touching the plaquette $p$ and acts trivially
everywhere else in the lattice (see Fig.~\ref{fig:stabilizers}). Similarly,
for each vertex (or site) in the lattice, define a {\em vertex} operator $S_v =
\bigotimes_{e \in \delta v} X_e$, where $\delta v$ denotes the edges
incident at vertex $v$ and $X_e$ is the Pauli $X$ operator acting on the
qubit at edge $e$.  In other words, $S_v$ acts as a tensor product of Pauli
$X$ operators on all the edges surrounding a vertex and acts trivially on
all the other qubits in the lattice, as shown in Fig.~\ref{fig:stabilizers}.

It is important to note that the rough and smooth boundaries still have
plaquette and vertex operators defined on them; these operators simply act
nontrivially on fewer qubits than the operators in the bulk of the lattice.
Since the lattice ${\mathcal L}$ has $l^2$ plaquettes and $l^2$ vertices,
there are also $l^2$ plaquette operators and $l^2$ vertex operators.  These
operators are all independent in the sense that no strict subset can
generate the rest, and, moreover, they all commute since they are incident
on each other an even number of times.

\begin{figure}[t]
\begin{center}
\includegraphics[width=0.95\columnwidth]{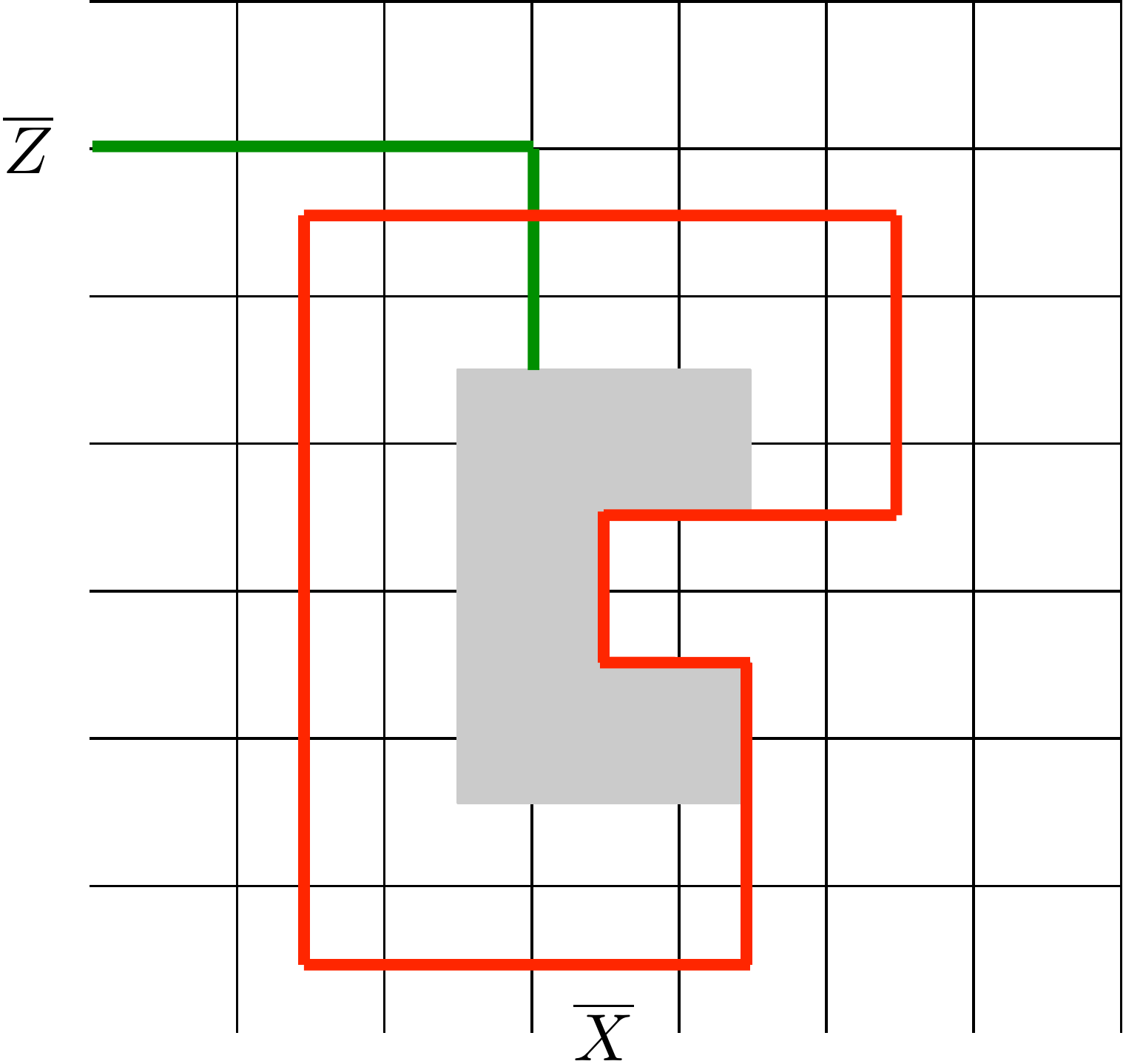}
\caption{A rough ($X$-type) defect.  A logical $X$ operator is defined by a
closed loop of $X$s on the dual lattice that surrounds the defect and a
logical $Z$ operator is defined by a connected path of $Z$s on the lattice
from the defect to a {\em rough} (\emph{$X$-type}) boundary. }
\label{fig:roughdefect}
\end{center}
\end{figure}

The collection of all the $S_p$ and $S_v$ operators comprises the set of
stabilizer generators for a quantum surface code, the codespace being
defined by the simultaneous $+1$ eigenspace of all the stabilizer
generators. This set generates the stabilizer group for the code, which is
simply the set of all the products of generators. The above description
actually specifies a single state rather than a codespace since it has
$2l^2$ checks on $2l^2$ qubits. This is a consequence of the particular way
in which we chose the boundary of the lattice, which disallows the existence
of any additional operators that commute with all of the generators but
which are not elements of the stabilizer group. Encoding quantum information
in the lattice requires the constructions described next.

Consider a closed simple curve $c$ on ${\mathcal L}$ that does not cross
itself and that does not touch the boundary of ${\mathcal L}$.  Call the
interior of this loop, excluding $c$ itself, $I_c$. Consider ``removing''
all of the qubits in $I_c$.  Here by ``removing'' we do not mean physically
removing the qubits, but rather that we consider a new code in which the
stabilizer generators exterior to the region $I_c$ are consistent with the
description above, while the region $I_c$ has a different set of stabilizer
generators (not necessarily of the plaquette and vertex type). We call this
process \emph{puncturing} (not to be confused with the notion of puncturing
associated with classical coding theory \cite{MacWilliams:1977a}), and the
resulting region of removed qubits is called a {\em defect}.  Given such a
defect, we can study the properties of the new code induced on the exterior
of $I_c$.  Careful counting of the stabilizer generators and qubits in this
new code reveals that the puncturing procedure has created a logical
qubit~\cite{Bravyi:1998a}. The logical operators for the new logical qubit
can be chosen as follows: an encoded $Z$ is a closed loop of $Z$ operators
on the lattice ${\mathcal L}$ that encircles the defect and an encoded $X$
is a connected path of $X$ operators on the dual lattice ${\mathcal L}^*$
that starts on the smooth ($Z$-type) boundary of the defect and ends on a
smooth ($Z$-type) boundary of the lattice ${\mathcal L}$ other than the
loop $c$ (see Fig.~\ref{fig:smoothdefect}). The distance of this code is the
minimum of the length of curves on ${\mathcal L}$ bounding the defect and
the length of paths connecting the defect to a smooth ($Z$-type) boundary of
${\mathcal L}$.  We note that the curve $c$ itself is the minimum-weight
choice for the encircling logical $Z$ operator. Similarly, instead of
starting with a simple closed curve on the lattice, we can consider a simple
closed curve on the dual lattice and remove the interior of this curve.  To
be consistent with the definition given for the former kind of defect, we
must define the encoded $X$ to be a closed loop $c^*$ of $X$ operators on
the dual lattice ${\mathcal L}^*$ that encircles the defect and the encoded
$Z$ to be a connected path of $Z$ operators on the lattice ${\mathcal L}$
that starts on the rough ($X$-type) boundary of the defect and ends on a
rough ($X$-type) boundary of the lattice ${\mathcal L}$ other than the
loop $c^*$ (see Fig.~\ref{fig:roughdefect}).

Puncturing the surface code creates a single encoded qubit.  By puncturing
multiple times we can create a code with more than one encoded qubit, one
for each additional puncture. The boundary curves of these defects can be on
the lattice, in which case we call the defect {\em smooth}
(\emph{$Z$-type}), or on the dual lattice, in which case we call the defect
{\em rough} (\emph{$X$-type}).  The distance of such a code is the minimum
of the distance between defects, the distance between a  defect and the boundary
of the lattice, and the circumference of a defect.  

Surface codes with defects were first explored within the framework of
active quantum error correction.  Here we consider an alternative situation
in which we construct a Hamiltonian with a ground space that is degenerate
and identical to the codespace of a quantum error correcting code. The
construction of such a Hamiltonian is easy from a theoretical point of view;
it is simply the negative sum of the stabilizer generators, ${\mathcal G}$,
\begin{equation}
H=-{\Delta \over 2} \sum_{S \in {\mathcal G}} S.\label{eq:ham}
\end{equation}
The constant in front is chosen so that all errors will have an energy
penalty of a least $\Delta$ (errors adjacent to a boundary will have this
penalty, while errors away from boundaries will have a penalty of
$2\Delta$). Since the set of generators is commutative, the eigenspaces of
$H$ can be labeled by their eigenvalues with respect to the operators $S$.
Because the eigenvalues of all the $S$ are $\pm 1$, the ground state of
this Hamiltonian is equivalent to the codespace of the quantum code
generated by $\mathcal{G}$: $S|\psi\rangle=|\psi\rangle$ for all $S \in
{\mathcal G}$.

Hamiltonians like that in Eq.~(\ref{eq:ham}), which we call \emph{stabilizer
Hamiltonians}, have interesting properties for protecting quantum
information.  The first property is that operators that act nontrivially on
the codespace (the degenerate ground space) must be nonlocal, having a
Pauli-weight at least as large as the code's distance.  This allows the
system to retain its information even when perturbed by a local
Hamiltonian~\cite{Hamma:2008a, Bravyi:2010a}.  For toric codes, surface
codes, color codes, and, more generally, codes formed from quantum double
models \cite{Propitius:1994a}, this is a partial indication of a topological
order in the system.  (A more robust indicator would be a nontrivial
topological entanglement entropy~\cite{Hamma:2005a,
Kitaev:2006a,Levin:2006a,Flammia:2009b}.)

While a stabilizer Hamiltonian is robust to local perturbations, if the
system is immersed in a thermal bath, the lifetime of information encoded
into the ground state does not necessarily scale with the size of the system
(or the size of the defect for a surface code with defects).  For example,
for the toric code, the lifetime of this information is proportional to
$\exp(2\beta \Delta)$~\cite{Alicki:2009a}, where $\beta=(k_B T)^{-1}$ is the
inverse temperature of the bath. It is widely believed that all stabilizer
Hamiltonians with local terms embedded in two spatial dimensions have a
similar lifetime~\cite{Bravyi:2008a}.  The more challenging issue is how to
compute with them without increasing the rate at which information is
destroyed.  As mentioned in Sec.~\ref{sec:introduction}, if a stabilizer
Hamiltonian describes a topologically ordered system possessing anyons with
a sufficiently rich nonabelian structure, then quantum computation can be
carried out by creating, braiding, and fusing the anyons.  However, it is
not entirely clear that one can controllably create single excitations
without also creating other uncontrolled excitations that could then
disorder the system, nor how one can move the anyons without causing other
anyons to be produced.  This has led to the search for self-correcting
quantum systems where the excitations are not point-like particles like
anyons but structures that have boundaries with
dimension~\cite{Dennis:2002a, Bacon:2006a, Bravyi:2008a}.  The energetic
cost of an excitation in such a system is proportional to the size of its
boundary and thus would be robust to errors during creation and movement
processes---such a system would energetically favor shrinking the boundaries
of the errors to zero, causing them to vanish.  In particular, it has been
argued that such systems would have a lifetime proportional to their size,
indicating that the system and the environment to which it is coupled
participate in a form of ``self-correction'' in which the environment that
creates the errors can also fix the errors; at a low enough temperature, the
rate of the latter process dominates the rate of the former.  In this paper,
we do not directly address the question of self-correction; instead we
attempt to better understand how computation can be done adiabatically
within existing models.

%%%%%%%%%%%%%%%%%%%%%%%%%%%%%%%%%%%%%%%%%%%%%%%%%%%%%%%%%%%%%%%%%%%%%%%%%%%%%%%
% Section
%
\section{Adiabatic code deformations}
\label{sec:agt}

Before showing how to perform the adiabatic deformations and creation of
fiducial states, we briefly review a scheme for performing adiabatic gate
teleportation~\cite{Bacon:2009a} (AGT), as this gives an idea of how the
protocols we introduce below operate.  AGT is a procedure for transferring
information in one qubit to information in another qubit (with a possible
gate applied to this information) via the use of an adiabatic evolution and
an ancillary qubit.   The following example is on a system composed of three
qubits in which the first and third qubit are swapped (without a gate
applied during the swapping).  Initially the system evolves under a
Hamiltonian given by
\begin{equation}
H_i=- \Delta(I_1 X_2 X_3 + I_1 Z_2 Z_3),
\end{equation}
where $P_i$ represents the operator $P$ acting on the $i$th qubit and where
we soon omit the identity operators $I$. A final Hamiltonian is defined as 
\begin{equation}
H_f=- \Delta(X_1 X_2 I_3 + Z_1 Z_2 I_3).
\end{equation}
The AGT protocol begins with the information encoded in the first qubit and
$H_i$ turned on.  Then, $H_i$ is adiabatically turned off while
simultaneously turning on $H_f$.  In other words, the evolution is described
by
\begin{equation}
H(t)=f(t)H_i + g(t) H_f, \label{eq:adiaevo}
\end{equation}
where $f(0)=1$, $f(T)=0$, $g(0)=0$ and $g(T)=1$ and $T$ is the time taken to
perform the evolution.  If $f(t)$ and $g(t)$ are chosen to be slowly varying
and the time $T$ is long enough such that the evolution is adiabatic
(meaning here that the probability of exciting the system out of its ground
space is made small), then the above evolution will take information in the
first qubit and send it to information in the third qubit.  For example, one
may choose $f(t)=1-g(t)$ and $g(t)={t \over T}$ so that the evolution is
made adiabatic for sufficiently large $T$. A constant error can be achieved
for a fixed constant $T$.  

To see that a constant energy gap is maintained during the above evolution
and that the information is transported from the first to third qubit, it is
convenient to use the formalism of stabilizer codes to describe this
evolution.  Indeed, it is actually useful to define three codes.  The first
code, call it $S_1$, is defined by the stabilizer generators $X_2 X_3$ and
$Z_2 Z_3$ and the logical Pauli operators $\overline{Z}=Z_1 Z_2 Z_3$ and
$\overline{X}=X_1 X_2 X_3$.  A second code, call it $S_2$, is defined by the
stabilizer generators $X_1 X_2$ and $Z_1 Z_2$ and the logical Pauli
operators $\overline{Z}=Z_1 Z_2 Z_3$ and $\overline{X}=X_1 X_2 X_3$.
Suppose information is encoded into the stabilizer code $S_1$ so that it is
in the $+1$ eigenstate of both $X_2 X_3$ and $Z_2 Z_3$.  Notice then that
because $X_1=\overline{X} (X_2 X_3)$ and $Z_1=\overline{Z} (Z_2 Z_3)$,
information encoded into this code can be accessed by making a measurement
on the first qubit.  Similarly, information encoded into the second code,
$S_2$, is localized in the third qubit.  The adiabatic evolution in
Eq.~(\ref{eq:adiaevo}) can now be seen as adiabatically dragging a
Hamiltonian that is a sum over stabilizer generators in $S_1$ to a sum over
stabilizer generators in $S_2$ such that the information in the encoded
qubit described by $\overline{X}$ and $\overline{Z}$ is not touched.    

To analyze how the dragging between $S_1$ and $S_2$ occurs, it is useful to
introduce a new code, $S_3$.  This code has no non-identity stabilizer
operators, but has three encoded qubits.  These are defined by
\begin{align}
\overline{X}_1&=X_1 X_2 & \overline{Z}_1&=Z_2 Z_3 \nonumber \\
\overline{X}_2&=X_2 X_3 & \overline{Z}_2&=Z_1 Z_2 \nonumber \\
\overline{X}_3&=X_1 X_2 X_3 & \overline{Z}_3&=Z_1 Z_2 Z_3 
\end{align}
Notice that $\overline{Z}_1$ and $\overline{X}_2$ are the stabilizer
generators of $S_1$ and $\overline{X}_1$ and $\overline{Z}_2$ are the
stabilizer generators for $S_2$.  From this perspective, then, the adiabatic
evolution is from the initial Hamiltonian $-\Delta
(\overline{Z}_1+\overline{X}_2)$ to the final Hamiltonian
$-\Delta(\overline{X}_1 + \overline{Z}_2)$.  These are then simple
interpolations between single operators on encoded qubits, and will have a
constant energy gap.  Indeed, both $S_1$ and $S_2$ can be turned into $S_3$
by promoting stabilizer generators in these codes to logical Pauli
operators.  When it is possible to perform such a change between codes via
an adiabatic evolution, we say that we can adiabatically {\em deform} one
code into the other.  This technique is at the heart of the constructions in
this paper.  

To see that the information encoded in the first qubit ends up at the third
qubit, first note that, during the above evolution, the third encoded qubit
is not involved.  This implies that information encoded into this qubit will
not be affected by the evolution.  Next note that
$X_1=\overline{X}_3\overline{X}_2$, $Z_1=\overline{Z}_3 \overline{Z}_1$ and
$X_3=\overline{X}_3 \overline{X}_1$, $Z_3=\overline{Z}_3 \overline{Z}_2$.
Recall that we are dragging between the $+1$ eigenstate of $\overline{X}_2$
and $\overline{Z}_1$ to the $+1$ eigenstate of $\overline{Z}_2$ and
$\overline{X}_1$.  Thus, since information encoded into the third qubit is
not changed during the above evolution, we see that the protocol transports
the information in the first qubit to the third qubit.

More generally, the AGT protocol can be extended to enable universal quantum
computation~\cite{Bacon:2009a}.  We omit the details of this construction
except for noting that even when generalized, the energy gap used to
guarantee adiabatic evolution is a {\em constant} with respect to the number
of qubits in the system.  We will often refer to this by saying that the
energy gap of an adiabatic evolution is {\em constant} when considered by
itself---we use this language merely to imply that stringing together
similar parallel evolutions will not shrink the gap as a function of the
number of qubits involved in the evolution.

%%%%%%%%%%%%%%%%%%%%%%%%%%%%%%%%%%%%%%%%%%%%%%%%%%%%%%%%%%%%%%%%%%%%%%%%%%%%%%%
% Section
%
\section{Adiabatic code deformations of the surface code}
\label{sec:acd}

With the punctured surface code defined, we now present a series of
adiabatic code deformations that allow for a nearly universal set of
operations. First, we show how to prepare a surface code without any
defects. Next, we show how to prepare smooth defects in the $+1$ eigenstate
of $\overline{Z}$ and rough defects in the $+1$ eigenstate of
$\overline{X}$. We then show how to prepare smooth defects in $\pm 1$
eigenstates of $\overline{X}$ and rough defects in $\pm 1$ eigenstates of
$\overline{Z}$. (These procedures prepare the defects in eigenstates of the
string-like logical operators that tether the defects to a boundary.)
Following this, we introduce a procedure to allow code regions containing
defects to be separated from and attached to the rest of the code. We next
show how defects can be deformed, allowing them to be moved around the
lattice. This additionally allows for the $CNOT$ to be enacted between a
smooth and a rough defect. Finally, we show how arbitrary ancilla states can
be injected into defects and utilized in a computation.

The procedures above can be performed in an entirely adiabatic fashion and
thus benefit from the protection of a Hamiltonian gap. Additionally,
procedures like defect braiding also benefit from the topological nature of
the surface code Hamiltonian, with logical errors requiring high-weight, 
correlated physical errors
corresponding to nontrivial cycles on the lattice or dual lattice. We
mention this now to highlight the difference between the entirely adiabatic
operations presented in this section and operations we present in
Sec.~\ref{sec:non_adiabatic}---such as measurement or heralded gate
application---that do not inherit any protection from the gap or the
topology.

%%%%%%%%%%%%%%%%%%%%%%%%%%%%%%%%%%%%%%%%%%%%%%%%%%%%%%%%%%%%%%%%%%%%%
% Subsection
%
\subsection{Creation of a surface code without defects}
\label{sec:surface_creation}

\begin{figure}[h]
\begin{center}
\includegraphics[width=1.0\columnwidth]{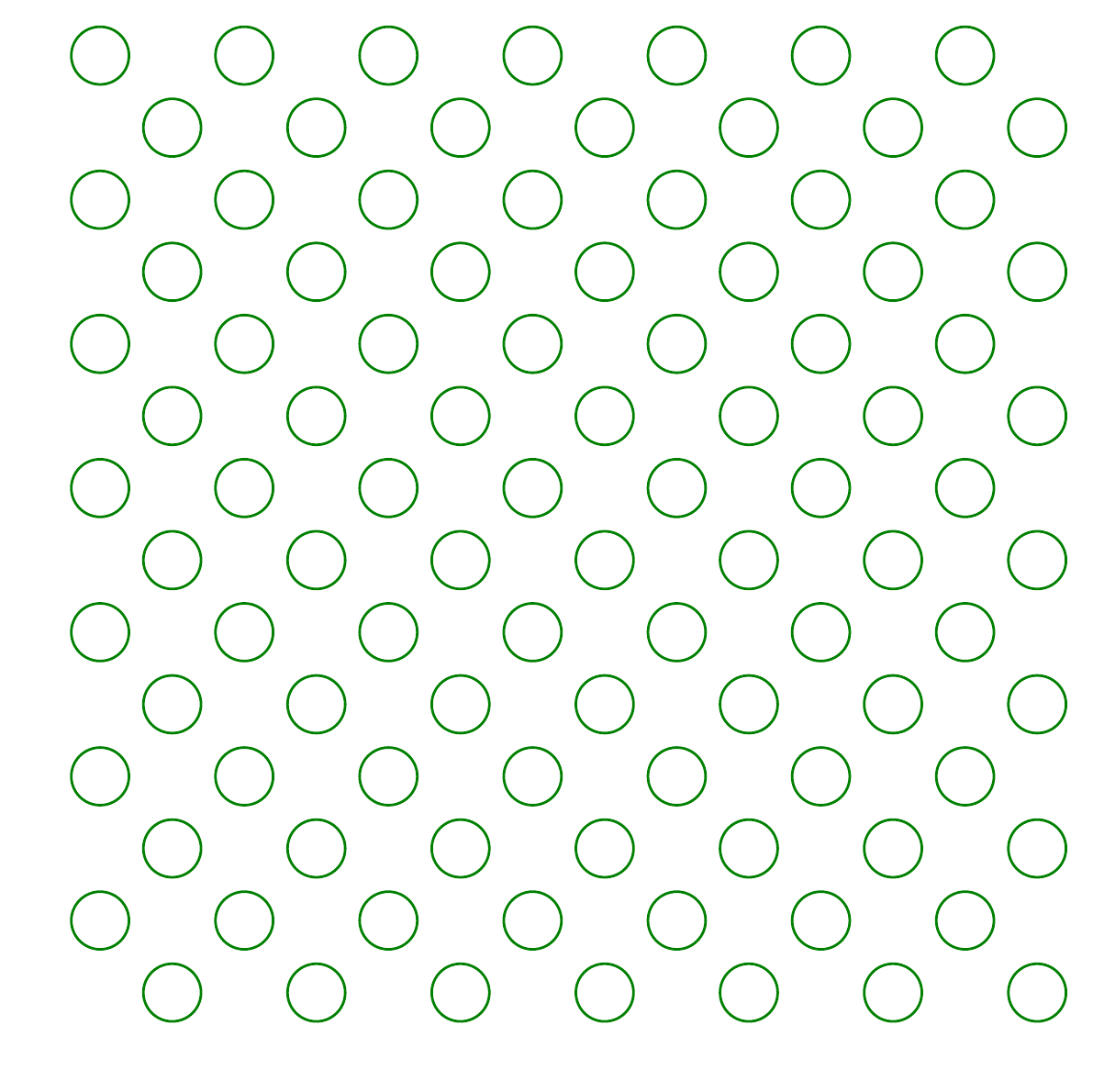}
\caption{A large array of qubits in the state $|0\>$, each protected by a
Hamiltonian $H = - \Delta Z$.\label{fig:surface_pre_create}}
\end{center}
\end{figure}

\begin{figure}[h]
\begin{center}
\includegraphics[width=1.0\columnwidth]{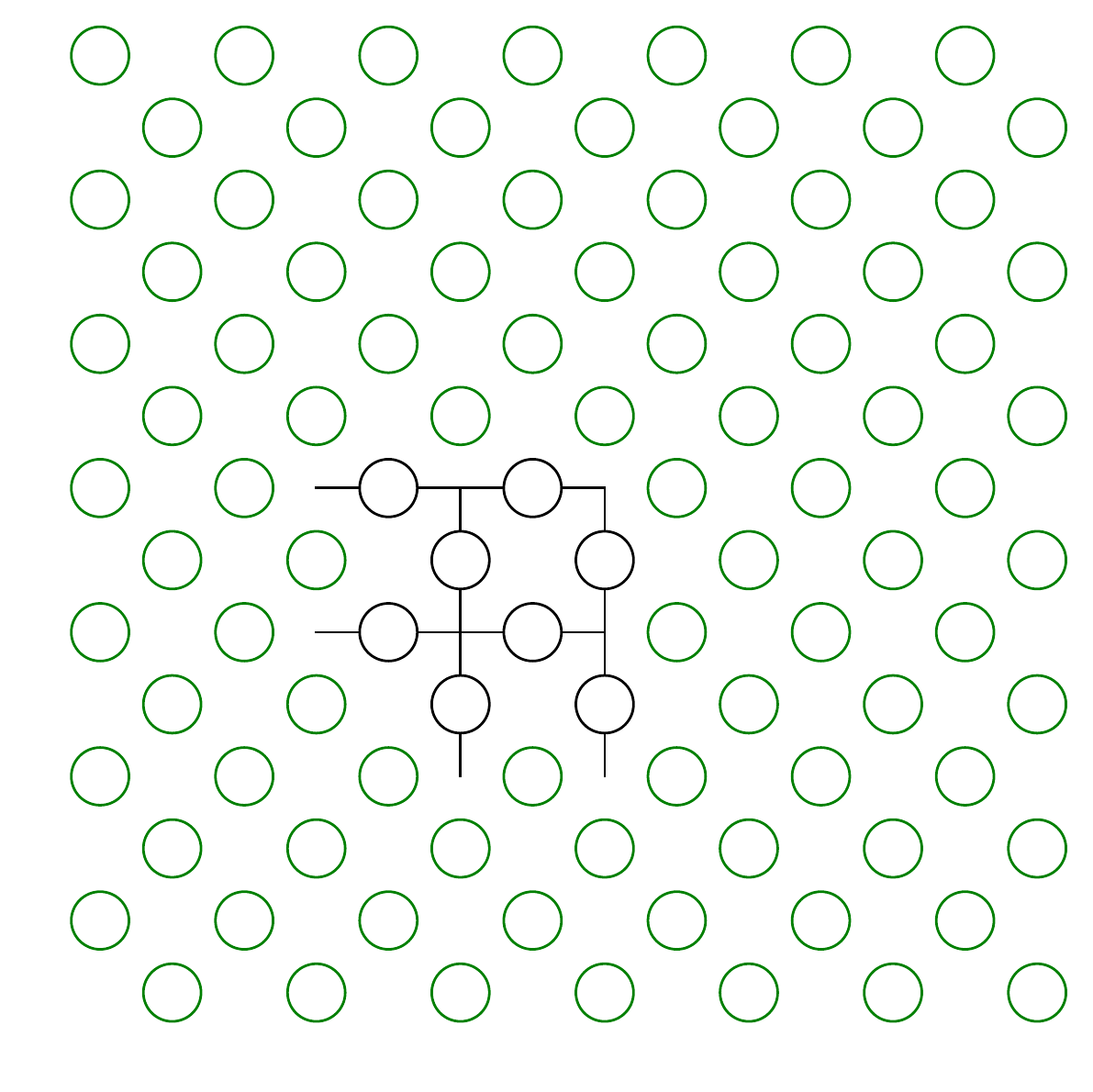}
\caption{A large array of qubits, an $8$-qubit region of which is now
encoded in the surface code (shown in black). The boundaries of the code are
chosen to be trivial so that the codespace is
nondegenerate.\label{fig:surface_create}}
\end{center}
\end{figure}

We begin by assuming that we have a large array of qubits, shown in
Fig.~\ref{fig:surface_pre_create}, stabilized by a Hamiltonian $H_i$ given
by \begin{equation} H_i = - \Delta \sum_j Z_j, \end{equation} where the sum
runs over all the qubits. The ground state of this Hamiltonian is unique and
has all the qubits in the state $|0\>$.
To prepare the surface, standard active error correction techniques call for
the stabilizer generators to be measured. Here, we simulate these
measurements in the vein of the ``forced measurements'' introduced in
Ref.~\cite{Bonderson:2013a} by slowly turning off $H_i$ and turning on the
Hamiltonian introduced in Eq.~(\ref{eq:ham}) for the specific instance of a
``small'' surface code. Turning on a Hamiltonian with a ``large'' surface
code as the ground state would cause the system gap to shrink proportionately
with the size of the code, so to be concrete we choose to evolve initially to
a Hamiltonian with a small surface code ground state. (We will subsequently
show how its size can be sequentially increased.) In other words, we
adiabatically follow the Hamiltonian
\begin{align}
\label{eq:surface_create}
\nonumber
H(t)
&= \left(1 - \frac{t}{T}\right) \sum_{j \in \mathcal{Q}}
      \left( - \Delta Z_j\right) 
  + \frac{t}{T} \sum_{S \in \mathcal{G}}
       \left(- \frac{\Delta}{2} S\right) \\ 
 & + \frac{t}{T} \sum_{j \not \in \mathcal{Q}} \left( - \Delta Z_j \right),
\end{align}
where $\mathcal{Q}$ is the set of qubits participating in the surface code
terms. In this case, $\mathcal{G}$ has $8$ elements, the four plaquette
operators and the four vertex operators shown in
Fig.~\ref{fig:surface_create}. Provided $T$ is large, the system will remain
in the ground state. As we showed before, the ground state of the
Hamiltonian in Eq.~(\ref{eq:ham}) is the codespace if a surface code. We
choose it to be nondegenerate by our choice of boundaries, although this is
not a necessity. After the evolution, the array of qubits looks like
Fig.~\ref{fig:surface_create}.

\begin{figure}[h]
\begin{center}
\includegraphics[width=1.0\columnwidth]{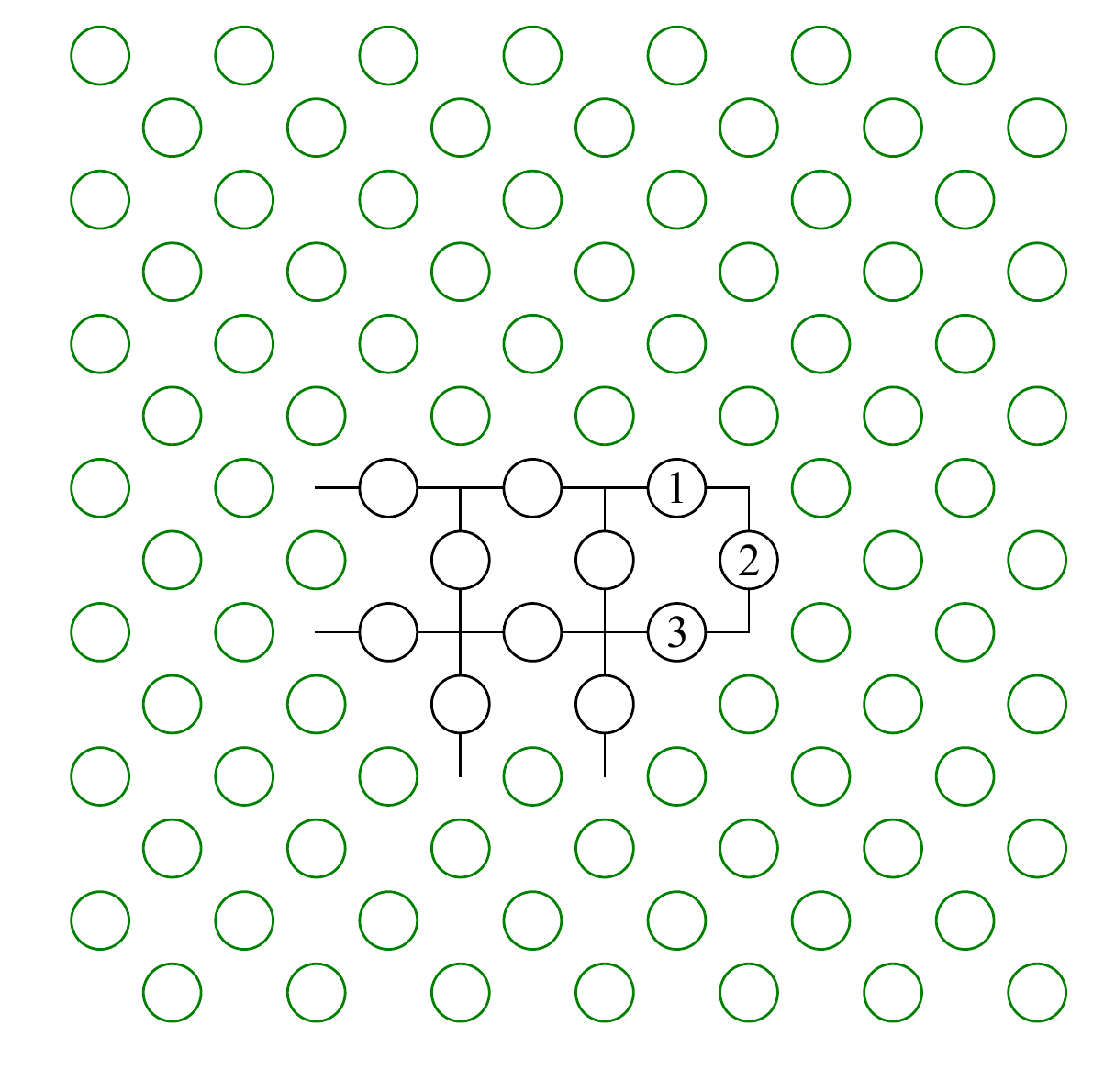}
\caption{Growth of a small surface code region that involves only the qubits
labeled $1$, $2$, and $3$.\label{fig:grow_small_surface}}
\end{center}
\end{figure}

Having created a small surface code that encodes no qubits, we can increase
its size by modifying the boundaries adiabatically. For example, we can grow
out part of the smooth boundary by performing an evolution of the form
\begin{align}
\nonumber
H(t)
 & = \left(1 - \frac{t}{T}\right)
            \left(- \Delta Z_1 - \Delta Z_2 - \Delta Z_3\right) \\
 &+ \frac{t}{T}\left(-\frac{\Delta}{2} Z_1 Z_2 Z_3
                     -\frac{\Delta}{2} X_1 X_2
                     -\frac{\Delta}{2} X_2 X_3\right),
\end{align}
where the numbering corresponds to Fig.~\ref{fig:grow_small_surface}.
This also requires the modification of vertex checks on the smooth
boundary being extended, which can be performed at the same time.
Additionally, a similar procedure will allow the extension of rough
boundaries. By piecing these additional evolutions together, a larger
surface code region can be constructed while maintaining a Hamiltonian gap
that is lower bounded by a constant proportional to $\Delta$.

For the remainder of this section, we will specialize our figures so that
they do not include the black dots that represent qubits, instead keeping
only the underlying square lattice structure of the code. However, the full
plane of qubits is still assumed to exist.

%%%%%%%%%%%%%%%%%%%%%%%%%%%%%%%%%%%%%%%%%%%%%%%%%%%%%%%%%%%%%%%%%%%%%
% Subsection
%
\subsection{Creation of a small \texorpdfstring{$Z$}{Z} (\texorpdfstring{$X$}{X}) defect in a \texorpdfstring{$+1$}{+1} eigenstate of
\texorpdfstring{$Z$}{Z} (\texorpdfstring{$X$}{X})}
\label{sec:create}

Here we describe how to create a two-plaquette smooth defect in an
unpunctured surface code. (The creation of a rough defect will proceed in an
exactly analogous way with the roles of $Z$ and $X$ interchanged.) We create
defects using two neighboring plaquettes for pedagogical clarity, although
creating single defects is also possible. With two-plaquette defects, it is
obvious that the creation process inherits protection from a Hamiltonian gap
and the topological nature of logical operators; for single-plaquette
defects, the Hamiltonian gap protection is not present.

\begin{figure}[t]
\begin{center}
\includegraphics[width=0.95\columnwidth]{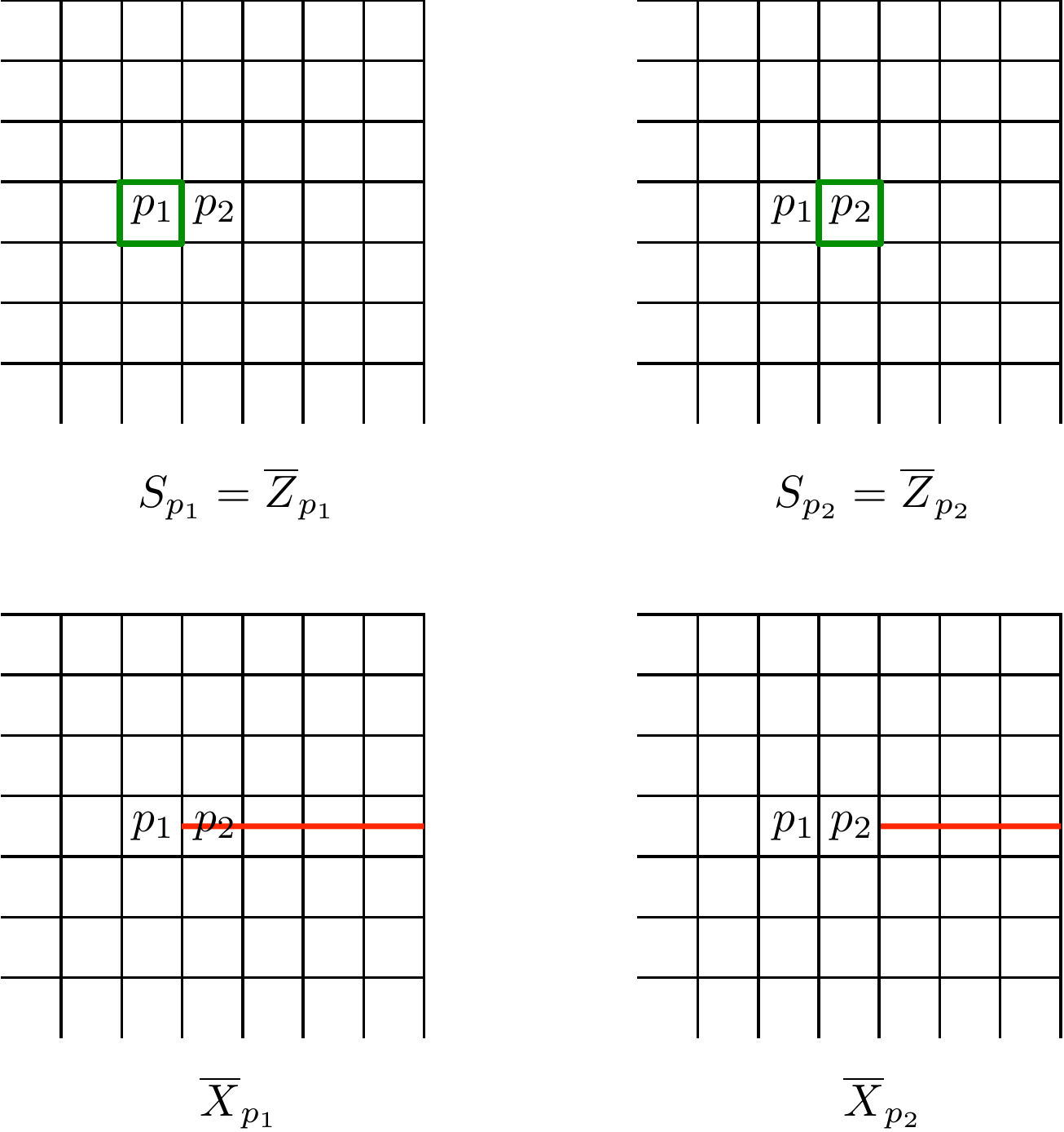}
\caption{Operators involved in creating the defect that includes $p_1$ and
$p_2$.  Note that the $X$ operations span to a nearby smooth boundary.}
\label{fig:create}
\end{center}
\end{figure}

To begin the creation procedure, the Hamiltonian is initially given by
Eq.~(\ref{eq:ham}), the negative sum of all the plaquette and vertex
stabilizer generators for the code. The defect will consist of two adjacent
plaquettes, bounded by a curve $c$ that encloses these plaquettes.  If the
stabilizer generators associated to these two plaquettes are $S_{p_1}$ and
$S_{p_2}$, we can promote them to $\overline{Z}$ operators for two encoded
qubits---$\overline{Z}_{p_1}$ and $\overline{Z}_{p_2}$ respectively---of a
new code where the stabilizer generators $S_{p_1}$ and $S_{p_2}$ have been
removed.  If we do this, then $\overline{X}$ for each qubit can be chosen as
a string of Pauli $X$ operators beginning on the appropriate plaquette,
traversing the dual lattice, and ending on a smooth boundary (see Figure
\ref{fig:create}).  In fact, we can always choose these operators so that
they overlap on all but the qubit separating the two plaquettes. We call
these two encoded logical $X$ operators $\overline{X}_{p_1}$ and
$\overline{X}_{p_2}$.  The operator $\overline{X}_{p_1}\overline{X}_{p_2}$
is then the single Pauli $X$ operator acting on the qubit between the
plaquettes.  

Suppose that we now perform the following adiabatic evolution: while turning
off the two plaquette operators, $S_{p_1}$ and $S_{p_2}$ in the Hamiltonian,
we simultaneously turn on the Pauli $X$ operator between these two
plaquettes.  In terms of the encoded logical operators we have defined
above, this is equivalent to starting with the Hamiltonian
\begin{equation}
H_i=-{\Delta \over 2} (S_{p_1}+S_{p_2}) = -{\Delta \over 2}
(\overline{Z}_{p_1}+\overline{Z}_{p_2}) \label{eq:createinitial}
\end{equation}  
and ending with the Hamiltonian
\begin{equation}
H_f=-{\Delta \over 2} \overline{X}_{p_1} \overline{X}_{p_2}
\label{eq:createfinal}
\end{equation}
All the other terms in the Hamiltonian commute with the relevant operators
and therefore do not contribute to any spectral shifts that might cause
crossings.  

In order to understand what happens in interpolating between $H_i$ and
$H_f$, it is convenient to note that $\overline{Z}_{p_1}\overline{Z}_{p_2}$
(which is a closed loop of Pauli $Z$ operators surrounding the smooth defect
we are creating) commutes with these Hamiltonians.  Also note that initially
the system is in the $+1$ eigenstate of both $\overline{Z}_{p_1}$ and
$\overline{Z}_{p_2}$, and hence also in the $+1$ eigenstate of
$\overline{Z}_{p_1}\overline{Z}_{p_2}$.  Because $\overline{Z}_{p_1}
\overline{Z}_{p_2}$ commutes with both $H_i$ and $H_f$, we may work in a
basis in which $\overline{Z}_{p_1} \overline{Z}_{p_2}$ and the full
Hamiltonian are simultaneously diagonal. This commutativity ensures that the
eigenvalue of $Z_{p_1} Z_{p_2}$ is conserved throughout the evolution. If we
perform this evolution via a simple adiabatic dragging between these
Hamiltonians (as described in Section \ref{sec:agt}) then the energy gap in
the system during this evolution remains constant. At the end of the
evolution, the system is in the $+1$ eigenstate of both $\overline{Z}_{p_1}
\overline{Z}_{p_2}$ and $\overline{X}_{p_1} \overline{X}_{p_2}$, which is
simply a single Pauli $X$ on the qubit between the plaquettes.

The above can be interpreted in terms of codes. By turning off two
stabilizer generators and turning on only a single Pauli $X$, we have
introduced an encoded qubit by decreasing the number generators. The product
of the two missing plaquette checks is $\overline{Z}$, and either
$\overline{X}_{p_1}$ or $\overline{X}_{p_2}$ can be chosen as
$\overline{X}$. Additionally, because the operator $\overline{Z}$ commuted
with the Hamiltonian throughout the adiabatic evolution, the encoded qubit
is prepared in the $+1$ eigenstate of $\overline{Z}$.

After this adiabatic evolution, the Hamiltonian does not quite factor into
two separate codes on the interior and exterior of the defect. The vertex
operators adjacent to the defect region still check the single qubit on the
interior. As a generating set, the four-body checks adjacent to the defect
and the single-body ``check'' on the interior qubit can equally well be
thought of as a generating set with two three-body operators that do not act
on the interior qubit, and the single-body operator that does. However, in
the Hamiltonian framework we must explicitly remove support of these
four-body checks on the interior qubits. We do this either by including the
modification of the adjacent vertex checks in the evolution discussed above,
or by using another evolution afterward that performs the modification. We
will assume that the former modification is used.

We note at this point that, while the defect we have created is small and thus
susceptible to relatively low-weight loops of $Z$ errors, these errors
actually have no effect. Since $\overline{Z}$ acts trivially on the state
we have prepared---namely, $|\overline{0}\>$---the fact that the defect has a
small perimeter is not detrimental. Once we start performing gates that
change the state, we will have to make sure that the perimeter is large, and
that the defect is far from the boundaries and other defects.

As mentioned above, the same arguments can be made for preparing rough
defects in the $+1$ eigenstate of $\overline{X}$. In that case, two
adjoining vertex checks are turned off while a single-body $Z$ on the qubit
in the middle is turned on. Two adjacent four-body $Z$ checks have to be
modified in this case, but the arguments are exactly the same as above.

It might be useful to address a question that may have entered the reader's
head. The procedures above adiabatically interpolate between a Hamiltonian
with a nondegenerate ground space to a Hamiltonian with a degenerate
ground space. Isn't there a level crossing between the ground space and an
excited space that can cause transitions away from the state we want to
prepare? Protection from this coupling is provided by the topological nature
of the logical operators. The only operator that can couple
$|\overline{0}\>$ and $|\overline{1}\>$ for a smooth defect is the
string-like operator $\overline{X}$ that connects the defect to a boundary.
This amounts to another way of saying that the eigenvalue of the operator
$\overline{Z}$ is a conserved quantity throughout the evolution, and so such
a crossing is not meaningful.

Now that we have introduced a method for creating smooth defects in the $+1$
eigenstate of $\overline{Z}$ and rough defects in the $+1$ eigenstate of
$\overline{X}$, we will show how these defects can be grown and moved
around the lattice. This will allow us to introduce other procedures, such
as the isolation of a defect from the bulk of the code and an adiabatic code
deformation that performs a $CNOT$ gate.

%%%%%%%%%%%%%%%%%%%%%%%%%%%%%%%%%%%%%%%%%%%%%%%%%%%%%%%%%%%%%%%%%%%%%
% Subsection
%
\subsection{Adiabatic deformation of defects}
\label{sec:deform}

We now show how to deform a defect. This involves modifying the Hamiltonian
by adding or removing stabilizer generators, the combination of which allows
defects to be moved. 

\begin{figure}[h]
\begin{center}
\includegraphics[width=0.95\columnwidth]{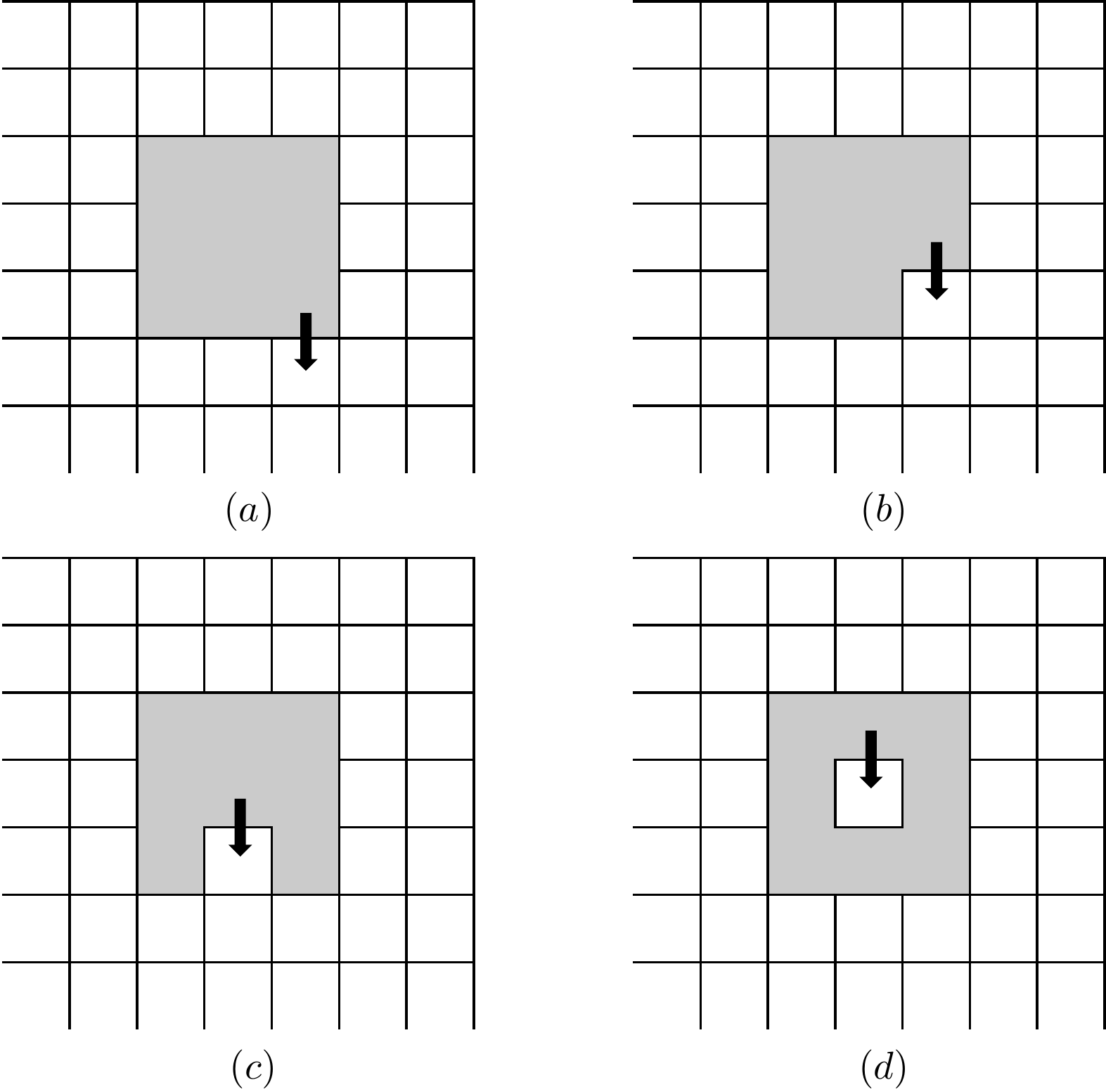}
\caption{The four potential situations faced when growing a smooth defect.
(a) Only one interior qubit. (b) Two interior qubits. (c) Three interior
qubits. (d) Four interior qubits.\label{fig:defect_interior_edges}}
\end{center}
\end{figure}

Consider a smooth defect that we wish to grow by turning off a single
adjacent plaquette check in the bulk of the system. The number of edges
bordering the interior of the defect is either $1$, $2$, $3$, or $4$, as
shown in Fig.~\ref{fig:defect_interior_edges}.
The procedure in each case is basically the same, with the clean-up or
potential removal of the adjacent vertex checks being the only difference.
The growth is achieved by turning off the plaquette check in the Hamiltonian
and turning on a single-qubit $-{\Delta \over 2}X$ Hamiltonian for each
qubit in the interior after the evolution. We also modify any adjacent
vertex checks at the same time to make the code factor properly into an
interior and an exterior.  We will briefly analyze the different interior
edge cases.

\begin{figure}[h]
\begin{center}
\includegraphics[width=0.95\columnwidth]{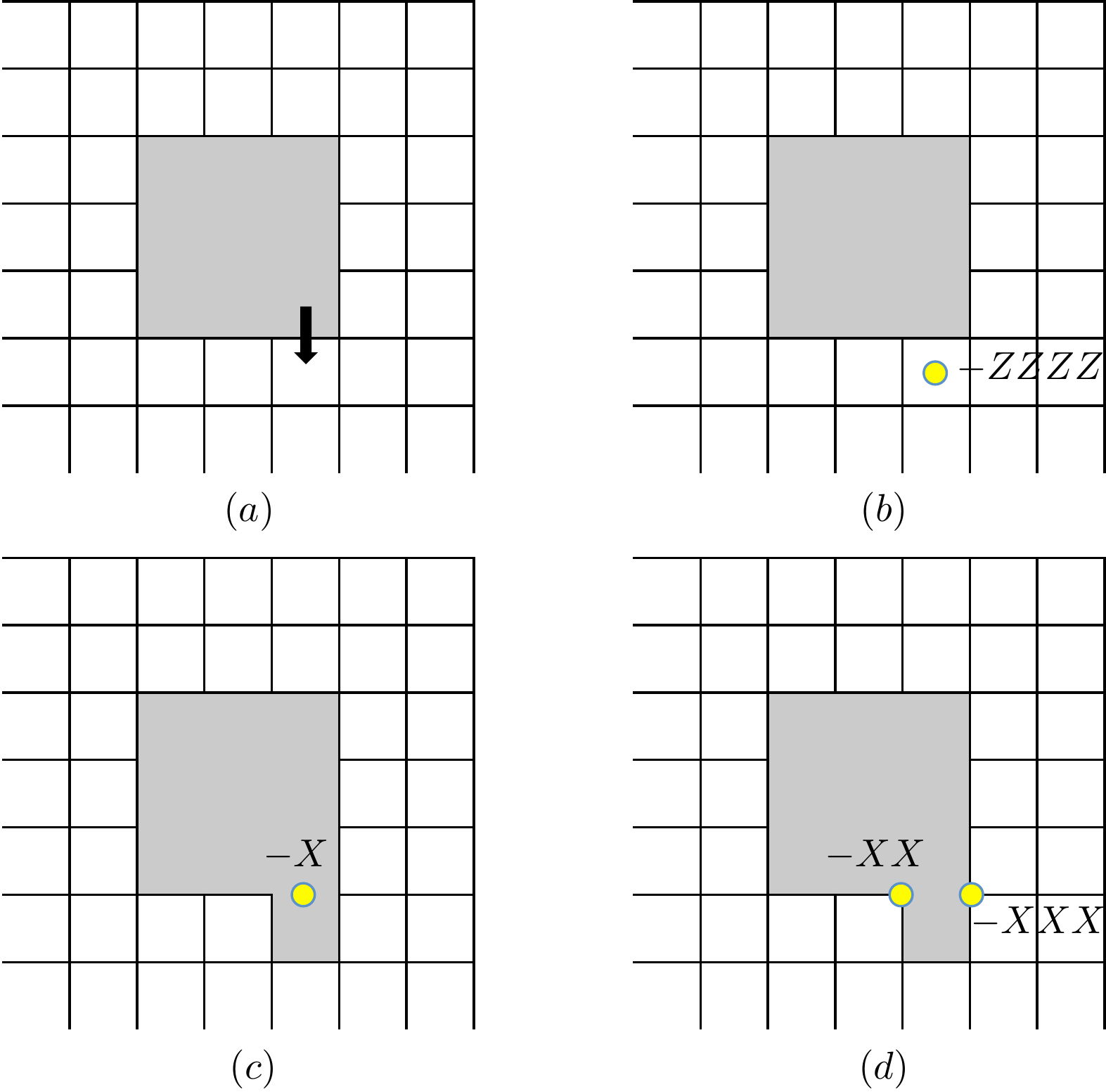}
\caption{Growth of a smooth defect with only a single qubit on the interior
after the procedure. (a) We wish to grow the defect to the indicated
plaquette. (b) We adiabatically turn off the neighboring plaquette while (c)
turning on a $-X$ Hamiltonian on the interior qubit. (d) This procedure
causes modifications to the neighboring $X$ checks that can be performed
simultaneously with steps (b) and (c).\label{fig:grow_one_edge}}
\end{center}
\end{figure}

For a single interior edge, as shown in Fig.~\ref{fig:grow_one_edge}, there
is not much different with respect to the case of defect creation. As the
plaquette check to grow into is turned off, a single-body $X$ on the qubit
adjacent to the defect and the plaquette is turned on. To fully sever the
interior and exterior regions, the only thing left to do is modify the two
adjacent vertex checks from three-body operators to two-body operators.

\begin{figure}[h]
\begin{center}
\includegraphics[width=0.95\columnwidth]{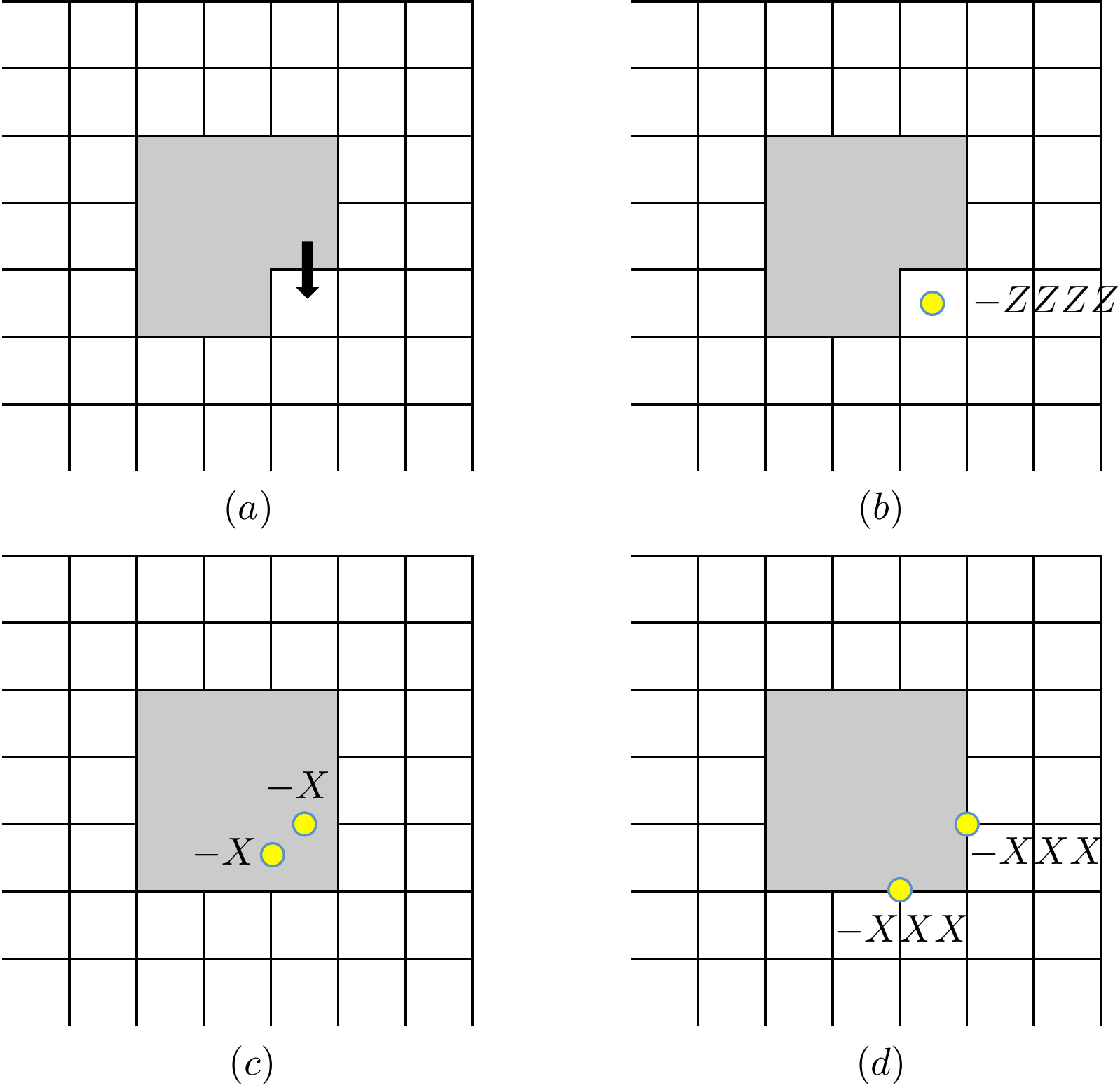}
\caption{Growth of a smooth defect with two qubits on the interior. (a) We
wish to grow the defect to the indicated plaquette. (b) We adiabatically
turn off the neighboring plaquette while (c) turning on two $-X$
Hamiltonians on the interior qubits. (d) This procedure causes modifications
to the neighboring $X$ checks.\label{fig:grow_two_edge}}
\end{center}
\end{figure}

\begin{figure}[h]
\begin{center}
\includegraphics[width=0.95\columnwidth]{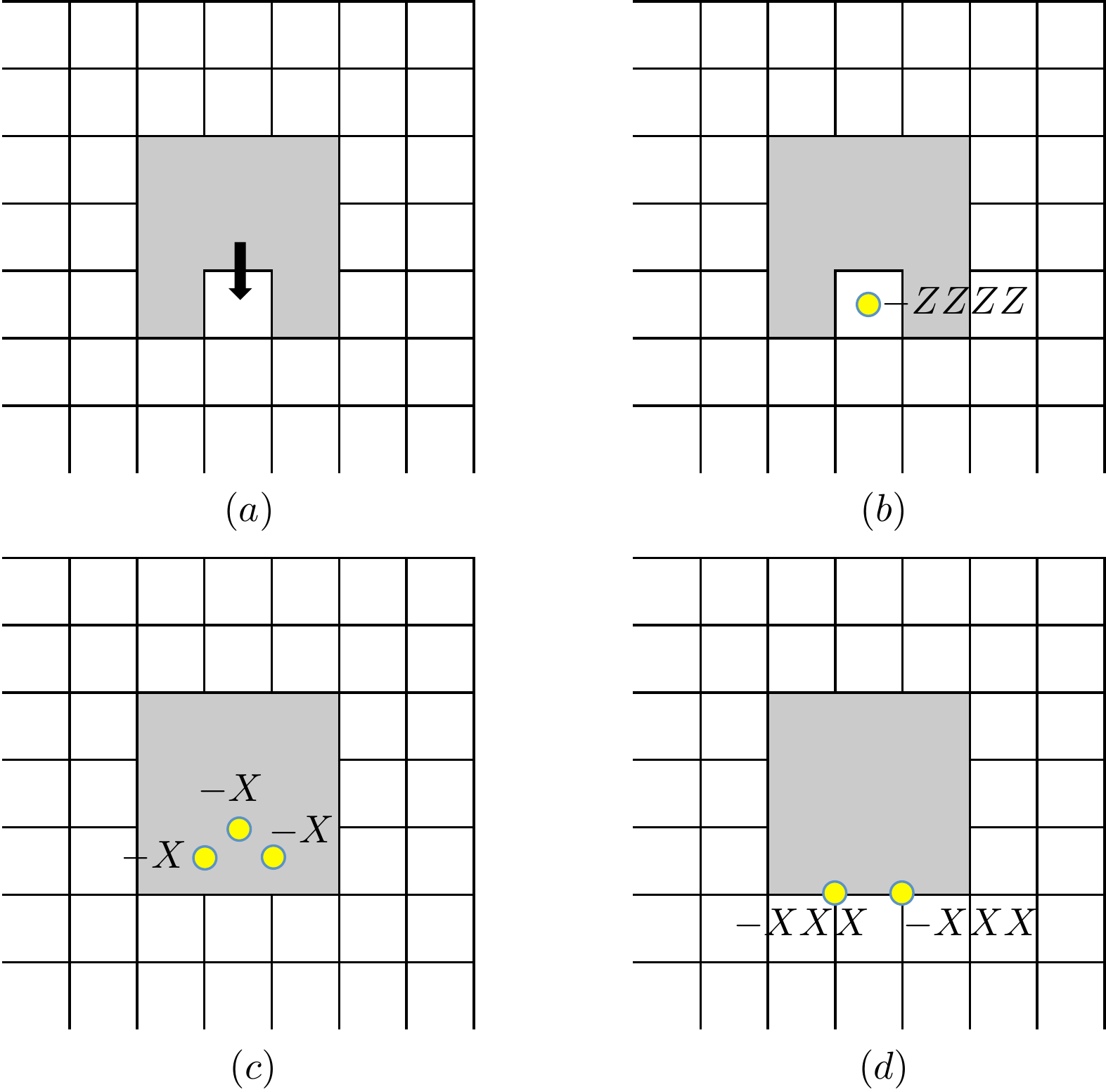}
\caption{Growth of a smooth defect with three qubits on the interior. The
process is essentially the same as the one depicted in
Fig.~\ref{fig:grow_two_edge}.\label{fig:grow_three_edge}}
\end{center}
\end{figure}

\begin{figure}[h]
\begin{center}
\includegraphics[width=0.95\columnwidth]{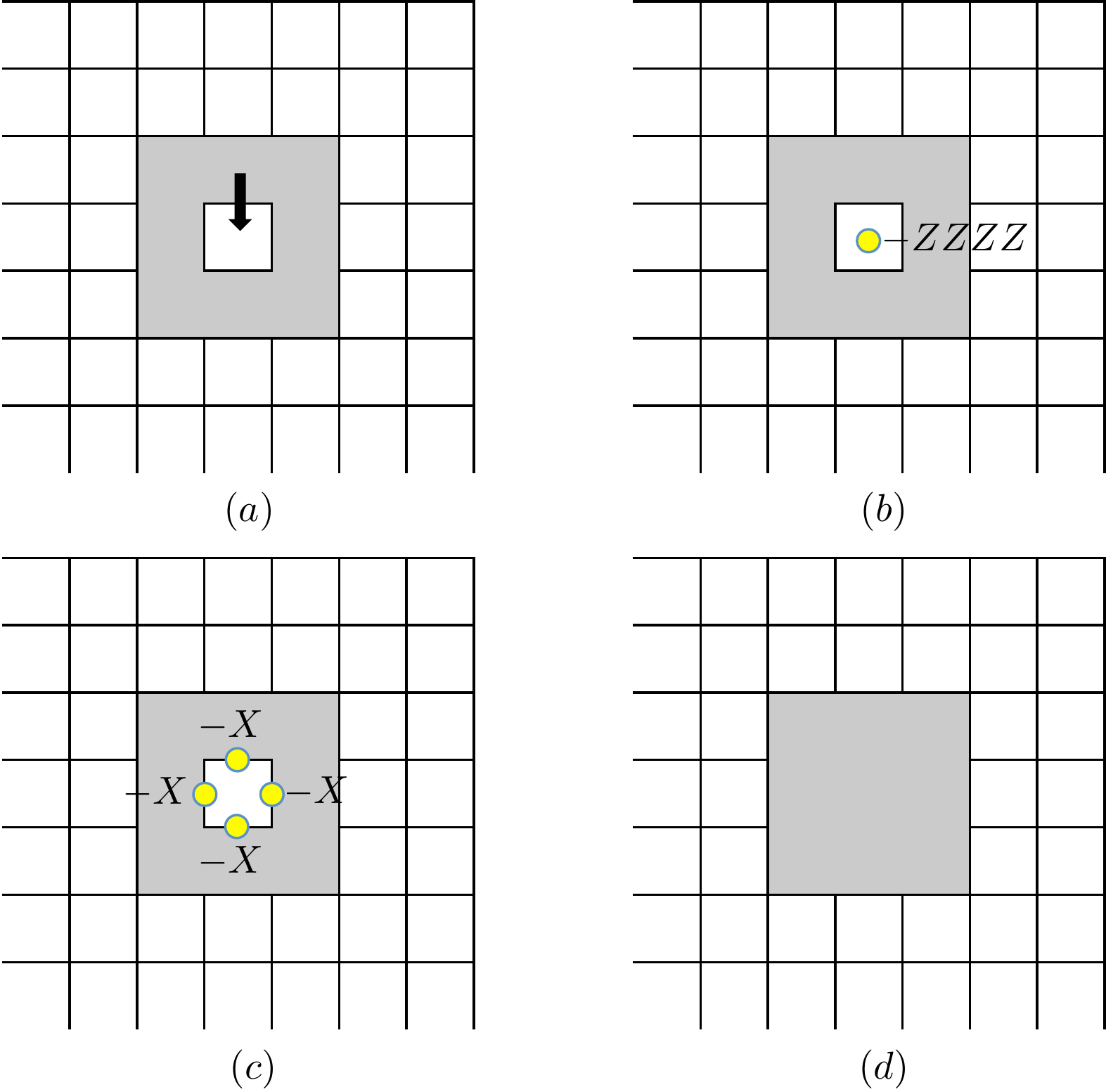}
\caption{Growth of a smooth defect with four qubits on the interior. The
procedure is the same as the others, but there are no resulting $X$ check
modifications.\label{fig:grow_four_edge}}
\end{center}
\end{figure}

The cases of $2$, $3$, and $4$ interior edges are different in that some
vertex checks are not only modified but are turned off completely. For the
case of $2$ interior edges, as shown in Fig.~\ref{fig:grow_two_edge}, the
appropriate evolution turns off the plaquette check while turning on two
single-body $X$ Hamiltonians on the interior edges. Note that the two-body
vertex check that operated on both the interior qubits is now redundant in
terms of stabilizer generators: it is simply the product of the two
single-body $X$ terms that were turned on. As such, it can simply be turned
off without having to worry about the codespace being affected; it merely
provides an additional energy penalty for errors on the two interior qubits.
The result is that we have removed two stabilizer generators---the plaquette
check and the two-body vertex check---and added two stabilizer
generators---the two single-body $X$ operators. Thus, we have not added any
additional logical qubits, we have merely grown the perimeter of an existing
one. As a final note, the two adjacent four-body vertex checks also must be
modified to three-body checks, and again, this can happen simultaneously
with the other adiabatic evolutions.
The case of $3$ and $4$ interior qubits, shown in
Fig.~\ref{fig:grow_three_edge} and Fig.~\ref{fig:grow_four_edge}
respectively, is almost identical. For the case of $3$, the plaquette check
is turned off while three single-body $X$ Hamiltonians are turned on. In
this case, two weight-two vertex checks are now redundant, and as before
they can simply be turned off without worrying about level crossings. The
counting works in a similar way, in that we have removed three stabilizer
generators and added three, preserving the number of logical qubits. The two
adjacent weight-four vertex checks also get modified to weight-three
operators. Finally, in the case of $4$ interior qubits, the same adiabatic
deformation is performed: the plaquette check is turned off and four
single-body $X$ Hamiltonians are turned on. Only three of the two-body
vertex checks are independent, and so only those three appeared in the
original Hamiltonian. They are the three checks made redundant by the
single-body $X$ Hamiltonians in this case. Unlike the other cases, in this
case there are no other vertex checks that need to be modified.

The procedure for shrinking defects is simply the inverse of the procedures
introduced above. By combining the ``grow'' and ``shrink'' operations, we
can move defects. As demonstrated in Ref.~\cite{Raussendorf:2007a}, an
encoded $CNOT$ gate can be performed by moving a smooth defect in a full
loop around a rough defect. The smooth defect is the control and the rough
defect is the target, and the direction of movement---clockwise or
counterclockwise---is unimportant.

%%%%%%%%%%%%%%%%%%%%%%%%%%%%%%%%%%%%%%%%%%%%%%%%%%%%%%%%%%%%%%%%%%%%%
% Subsection
%
\subsection{Detaching and attaching surface code regions with defects}
\label{sec:isolate}

\begin{figure}[t]
\begin{center}
\includegraphics[width=0.95\columnwidth]{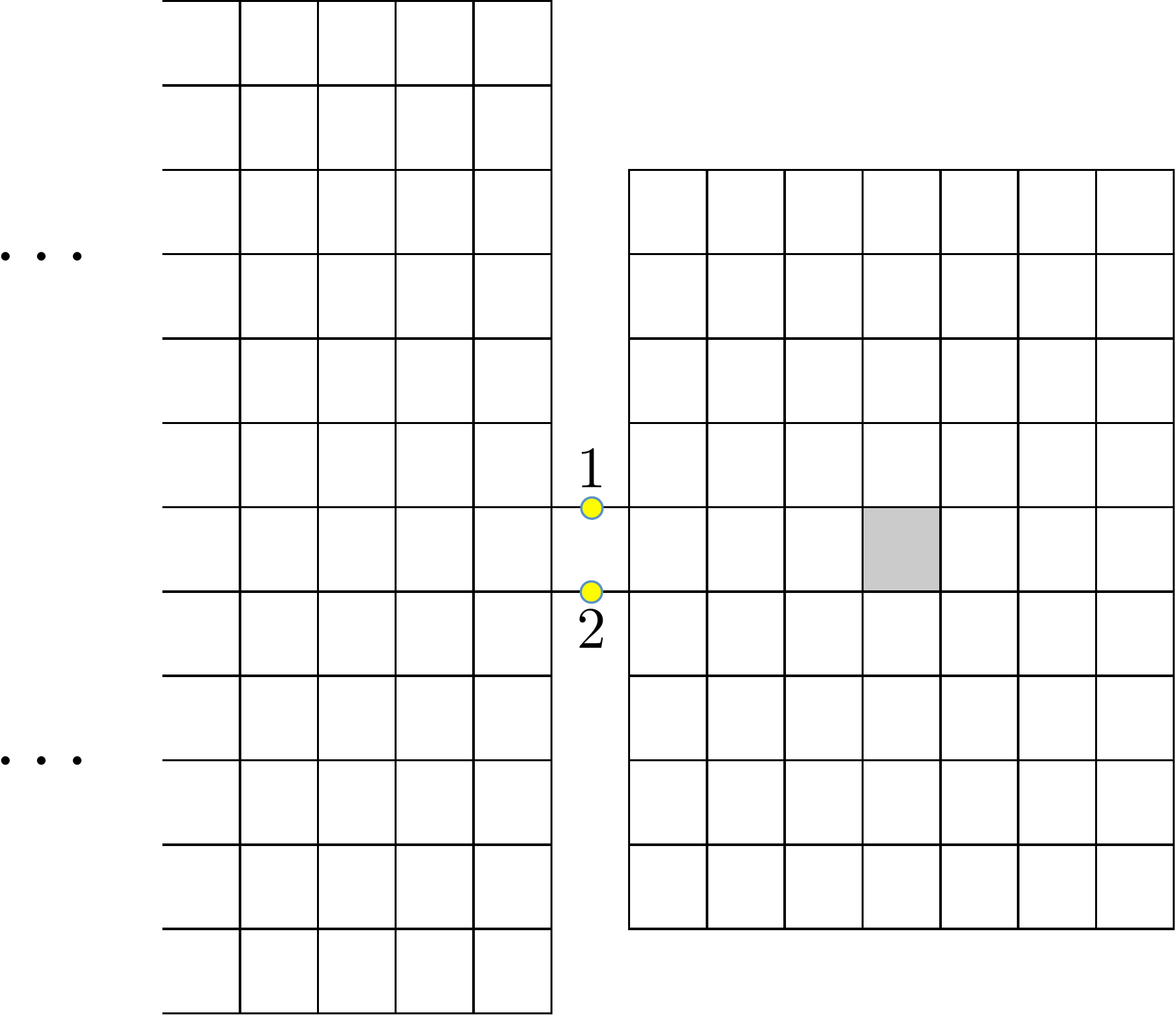}
\caption{The setup for pinching off a smooth defect from a smooth wall.}
\label{fig:pinch}
\end{center}
\end{figure}

For some subsequent procedures we will consider, it is helpful to have an
operation that isolates a defect from the surface code or reintroduces a
defect to the surface code that was previously isolated. By using defect
creation and growth operations described in Secs.~\ref{sec:create} and
\ref{sec:deform}, we can grow a defect ``moat'' around a defect of interest
so that the ``castle'' surrounding the defect has just a single
``drawbridge'' connecting it to the rest of the surface, as depicted in
Fig.~\ref{fig:pinch}.
The only additional operation we must consider to complete the isolation
procedure is how to ``lift the drawbridge'' by modifying the remaining check
operators adjacent to it. As before, we will only consider the case of
manipulating smooth defects---the case for rough defects is similar.

To isolate smooth defect, we must use smooth boundaries on the ``castle'' to
ensure that $\overline{X}$ for the defect will have a place to terminate
once the ``drawbridge'' is lifted. For concreteness, we assume that this
smooth boundary corresponds to the large boundary of the surface, but the
same procedure could be performed using a defect to create the isolated
region.

To remove the ``drawbridge,'' we simply turn off the single plaquette check
that connects the two regions while turning on a single-body $X$ on each of
the two qubits that need to be removed. (These qubits are labeled $1$ and
$2$ in Fig.~\ref{fig:pinch}.) The operator $X_1 X_2$, which was an element
of the stabilizer group before the evolution, is now redundant, just as in
the case of the interior checks that appear during defect growth in
Sec.~\ref{sec:deform}, and it is also removed. Thus, we remove two
checks---the check associated with the ``drawbridge'' and the two-body check
$X_1 X_2$---and replace them with two single-body $X$ checks in the
Hamiltonian. As before, the vertex checks adjacent to the ``drawbridge''
must be modified, and in this case they become three-body operators. (As a
closing aside, if we had tried to detach a smooth defect through a rough
boundary, the operator $X_1 X_2$ would no longer have been an element of the
stabilizer group.)

Reversing the detachment procedure allows regions with defects to be
attached to to the surface, introducing (or reintroducing) isolated defects
back into the code. This attachment procedure is an important step in our
protocols for making measurements of $\overline{X}$ and $\overline{Z}$ and
injecting ancilla states into the system, as discussed in
Sec.~\ref{sec:pauli_meas} and Sec.~\ref{sec:magic}. 
It is also possible to isolate and reintroduce a rough defect
through a rough boundary in an analogous fashion.

%%%%%%%%%%%%%%%%%%%%%%%%%%%%%%%%%%%%%%%%%%%%%%%%%%%%%%%%%%%%%%%%%%%%%
% Subsection
%
\subsection{Creation of a \texorpdfstring{$X$}{X} (\texorpdfstring{$Z$}{Z}) defect in a \texorpdfstring{$\pm 1$}{+/-1} eigenstate of
\texorpdfstring{$Z$}{Z} (\texorpdfstring{$X$}{X})}
\label{sec:esprep}

Another capability that will be useful for later procedures is the ability
to prepare rough defects in an eigenstate of $\overline{Z}$ and smooth
defects in an eigenstate of $\overline{X}$. The preparation of these defects in
performed in a region that is disconnected from the main surface.  It is
then attached to the surface using the procedure described in
Sec.~\ref{sec:isolate} to introduce it to the bulk surface.

To prepare a rough qubit in the $+1$ eigenstate of $\overline{Z}$, we
utilize a procedure very similar to the original creation of the surface,
described in Sec.~\ref{sec:surface_creation}. Recall that the stabilizer
Hamiltonian on a region disconnected from the surface is simply a sum of
single-body $-Z$ operators on each qubit. Once the location and size of the
disconnected region is chosen, we prepare it in a surface with solely rough
boundaries. Rather than following this up with the creation of a rough
defect, we simply prepare the surface by leaving a region of adjacent $X$
checks turned off and the single-body $Z$ terms on the interior of the
region unchanged. Since the system began in an eigenstate of any product of
$Z$ operators, and since $\overline{Z}$ for the rough qubit commutes with
all of the check operators we turn on, the system remains in the $+1$
eigenstate of $\overline{Z}$ after the evolution.

We also could have prepared the rough defect in the $-1$ eigenstate of
$\overline{Z}$ by first performing an adiabatic evolution on each qubit of
the form $-Z \rightarrow X \rightarrow Z$. This has the effect of dragging
each of the qubits into the $-1$ eigenstate of the local $Z$ operators, and
now, given a region of appropriate size, $\overline{Z}$ will have an
eigenvalue of $-1$ both before and after the defect creation process. (The
size constraints amount to ensuring that the weight of the logical operator
is odd.)

Smooth defects can be prepared in $\pm 1$ eigenstates of $\overline{X}$ in
much the same way, requiring only simple modifications. To prepare a smooth
defect in the $+1$ eigenstate of $\overline{X}$, each qubit first undergoes
the evolution induced by the adiabatic sequence $-Z \rightarrow -X$.
Likewise, to prepare a smooth defect in the $-1$ eigenstate of
$\overline{X}$, each qubit first undergoes the adiabatic evolution $-Z
\rightarrow X$. Now $\overline{X}$ will have the correct value before and
after the evolution that creates the defect, subject to the same size
constraints mentioned above.

%%%%%%%%%%%%%%%%%%%%%%%%%%%%%%%%%%%%%%%%%%%%%%%%%%%%%%%%%%%%%%%%%%%%%
% Subsection
%
\subsection{State injection into defects}
\label{sec:magic}

Creating defects in known ancilla states is another important building block
for our model. In typical architectures based on the surface code,
completing a universal set of encoded quantum gates requires the ability to
``distill'' high fidelity states---called ``magic states''---using protocols
like the one discovered by Bravyi and Kitaev \cite{Bravyi:2005a}. In this
section, we describe how to implement these preparations in an adiabatic
simulation of the process of state injection.

In measurement-based injection of a magic state \cite{Raussendorf:2007b},
one first exposes a qubit by preparing a single (unencoded) qubit in the
state $|\psi\>$. Then, the state is quickly encoded in a surface code
defect, and the procedure is finished by growing the defect to a
sufficiently large size so that it is well protected from noise. This
process need not be perfect, but any error introduced by the injection
procedure must keep the total error in the encoded state
$|\overline{\psi}\>$ below the threshold of the distillation protocol.

\begin{figure}[!htb]
\centering
\includegraphics[width=0.95\columnwidth]{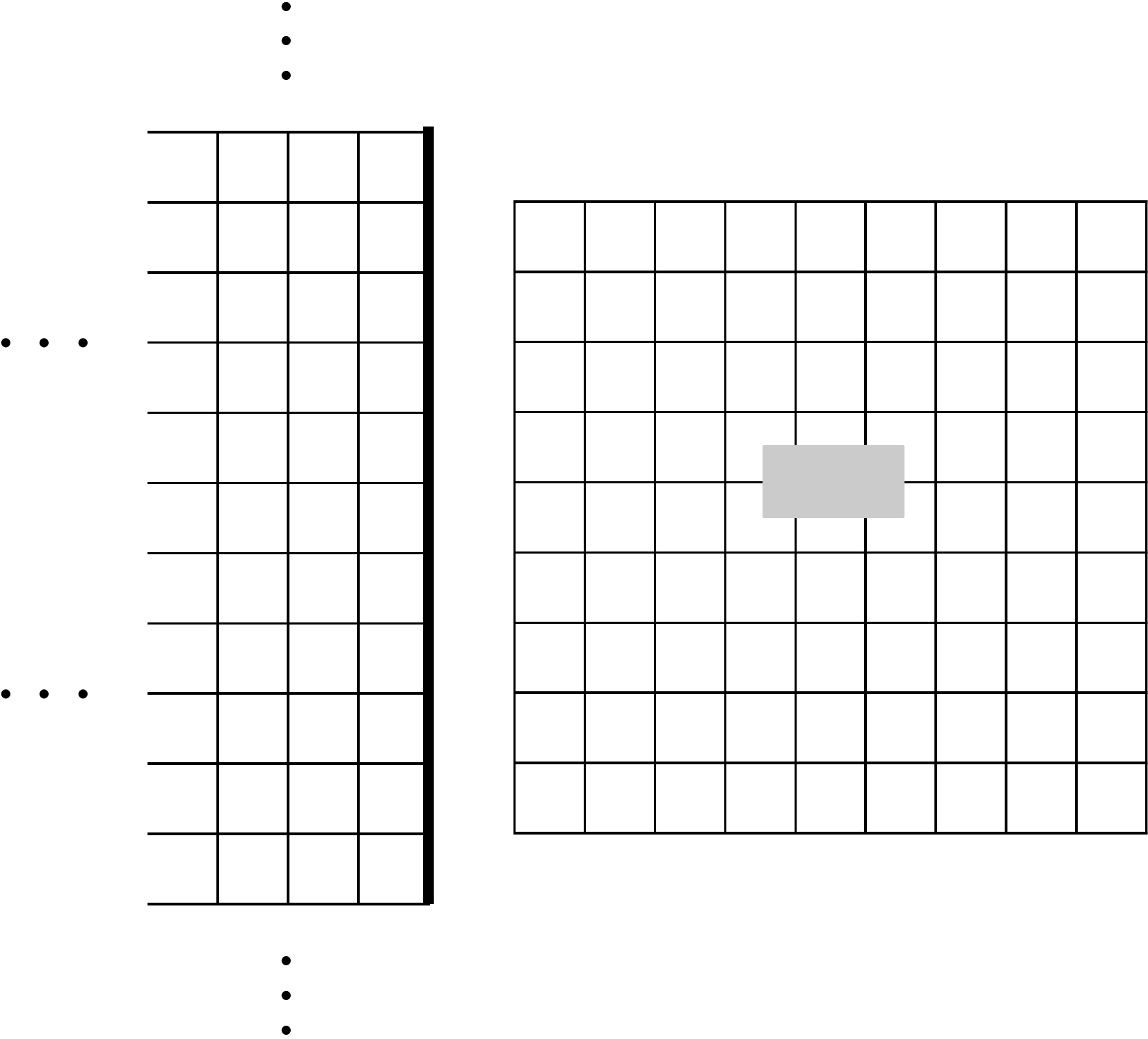}
\caption{After the Hamiltonian deformation (or sequence of deformations), we
are left with a surface code with trivial boundaries encoding a rough defect
in the state $\ket{+}$.\label{fig:magicprep2}}
\end{figure}

We describe our adiabatic simulation of this process for an injection into a
smooth defect, but the rough-defect case is similar. We begin by preparing
an all-smooth-boundary surface near the edge of the bulk surface using the
method described in Sec.~\ref{sec:surface_creation}. We then create a {\em
rough} defect in a $+1$ eigenstate of $\overline{X}$ in this region using
the procedure described in Sec.~\ref{sec:create}. The situation is depicted
in Fig.~\ref{fig:magicprep2}.
Because this region has only smooth boundaries, there is nowhere for a
string of $X$ operators from the defect to connect. Indeed, if we ignore the
one qubit on the interior of the defect, then what we would normally call
$\overline{X}$, a string of $X$ operators enclosing the defect, is already
an element of the stabilizer group. It can be formed by taking the product
of all the vertex checks. (As an aside, we note that this is a consequence
of the topology of the sphere, for which all loops remain homotopic when a
single point is removed.) As discussed in Sec.~\ref{sec:create}, this leaves
the single qubit on the interior of the defect in the $+1$ eigenstate of
$Z$.

\begin{figure}[!htb]
\centering
\includegraphics[width=0.95\columnwidth]{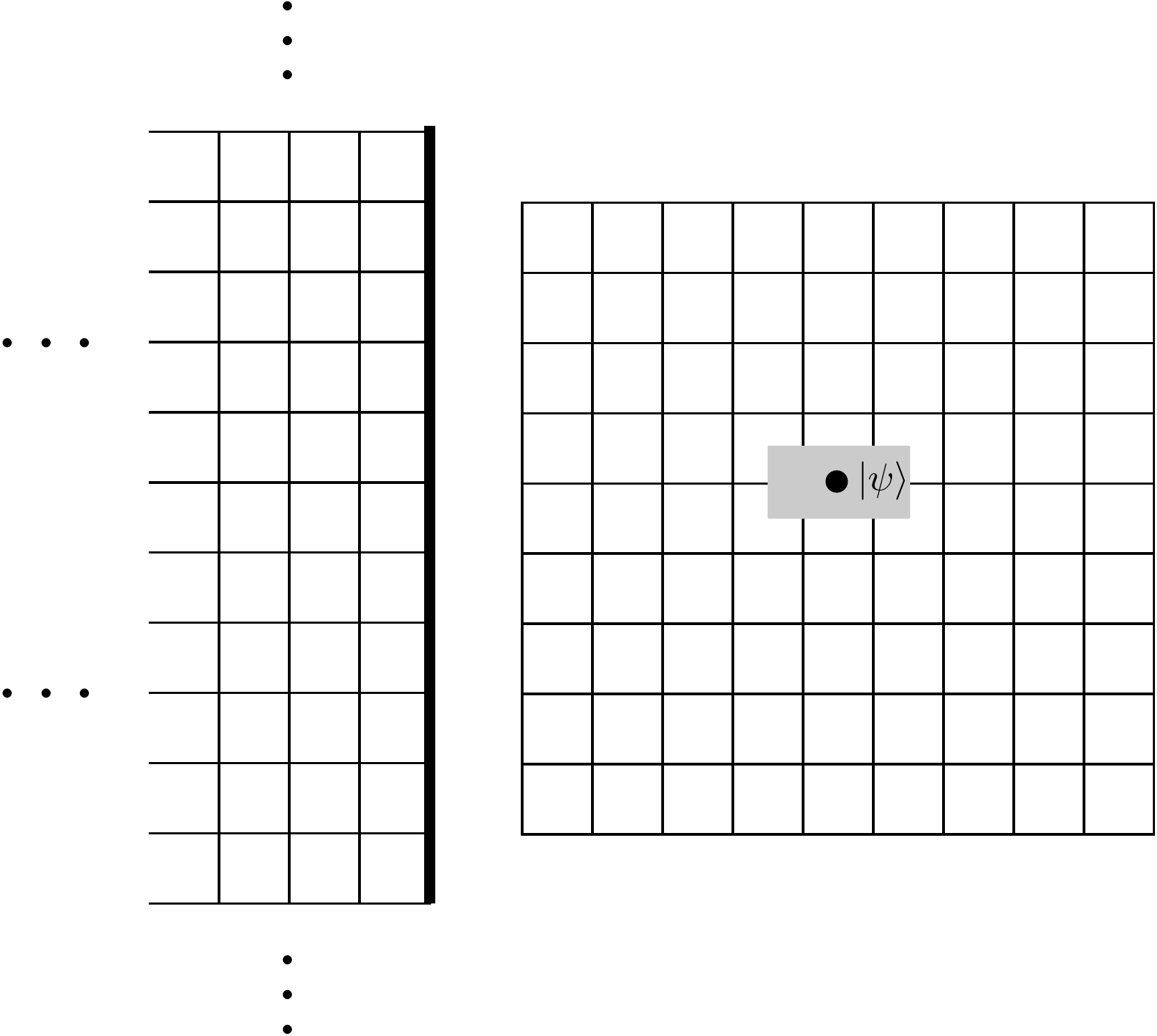}
\caption{The interior qubit is adiabatically dragged to the state $|\psi\>$,
the desired magic state.\label{fig:magicprep4}}
\end{figure}

\begin{figure}[!htb]
\centering
\includegraphics[width=0.95\columnwidth]{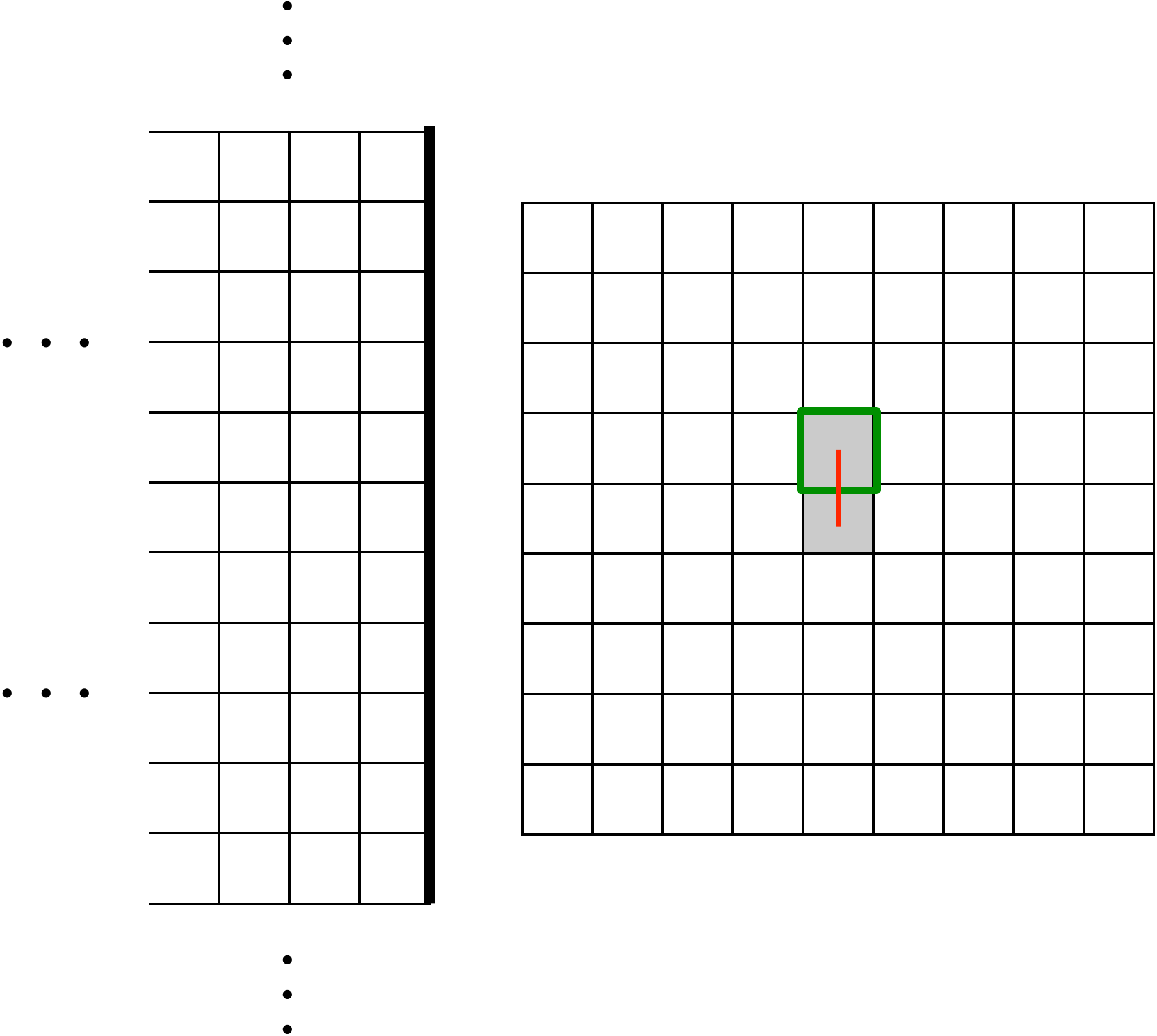}
\caption{The missing $X$ checks are reintroduced to the code, causing
neighboring $Z$ checks to be removed. This new defect is now encoded in the
state $|\psi\>$ with an encircling $\overline{Z}$ and a single-qubit
$\overline{X}$.\label{fig:magicprep5}}
\end{figure}

\begin{figure}[!htb]
\centering
\includegraphics[width=0.95\columnwidth]{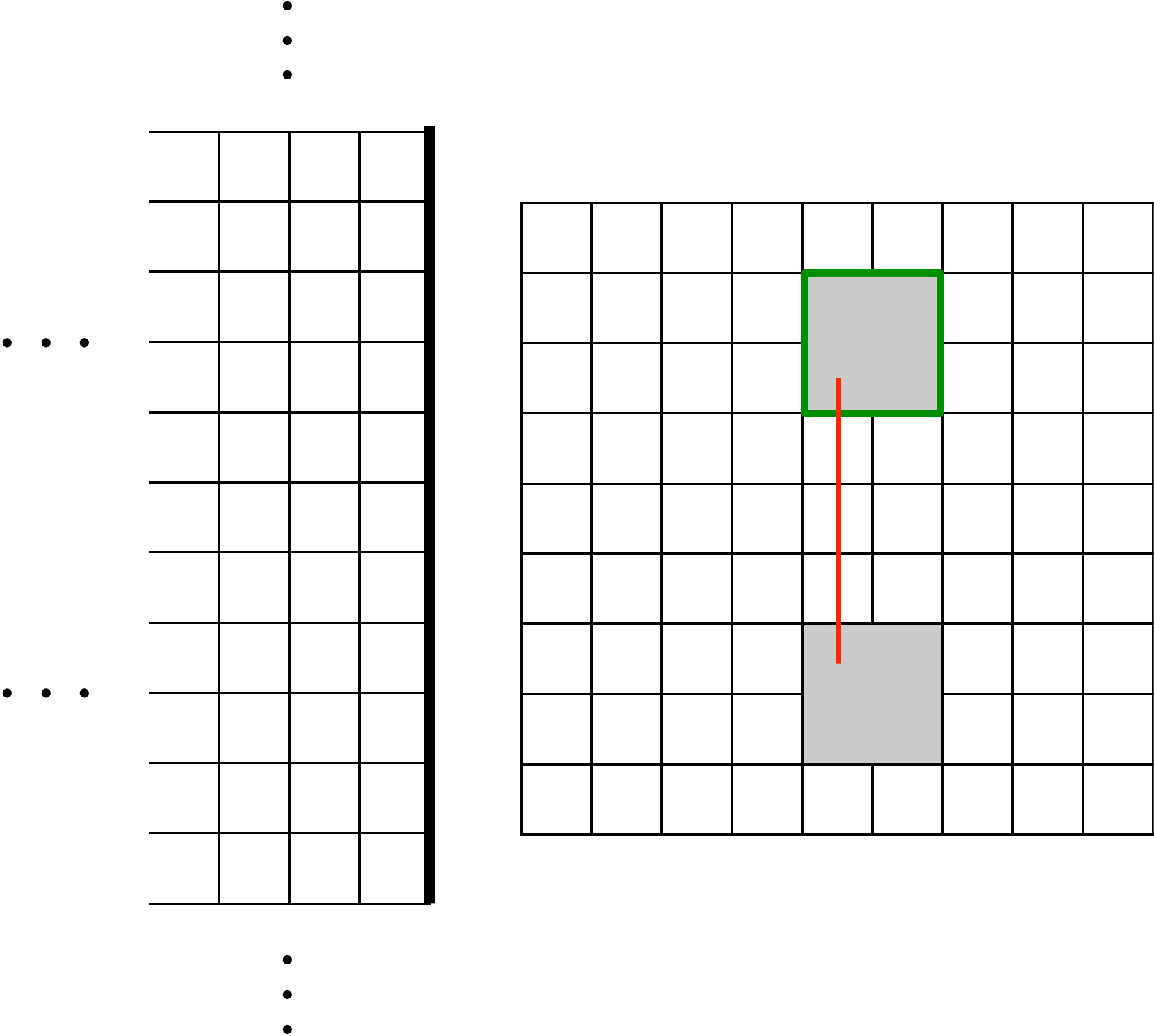}
\caption{Because $\overline{X}$ was only a single-qubit operator, the two
removed faces are moved apart and grown to combat
decoherence.\label{fig:magicprep6}}
\end{figure}

We then transform this interior qubit to the desired state by an adiabatic
evolution. For example, if we want to prepare the state $T|+\>$, we evolve
using the Hamiltonian $H(s) = (1-s)(-Z) + sUZU^\dagger$, where in this case
$U = TH$. If we think of this as a logical qubit, then $\overline{X}$ is a
single $X$ on the qubit and $\overline{Z}$ is a single $Z$ on the qubit.
Recall that the face checks originally incident on the interior qubit have
been modified and are no longer incident. The situation is now described by
Fig.~\ref{fig:magicprep4}.
Next, we adiabatically turn on the two vertex checks that were originally
turned off to create the defect. We simultaneously (and adiabatically) also
turn off the three-body plaquette checks, as they would otherwise
anti-commute with the final Hamiltonian. This evolution transforms the
logical operators, since the initial single-body $\overline{Z}$ does not
commute with the final $X$ checks. The transformation $\overline{Z}$
undergoes is determined by the Pauli algebra and the demands of a stabilizer
code. Since $\overline{Z}$ must still commute with the code after the vertex
checks are turned back on (note that the formerly interior qubit has now
been reintroduced to the code because the vertex checks are incident on it
once again), and since it also must not be in the stabilizer group itself, a
suitable choice of the new $\overline{Z}$ is the product of the old
$\overline{Z}$ and one of the three-body plaquette checks that also did not
commute with the vertex checks. What remains is what appears to be a normal
two-plaquette defect as shown in Fig.~\ref{fig:magicprep5}, but the crucial
difference is that there is now no sense of an isolated interior, since the
neighboring vertex checks are still incident on the qubit inside.
In fact, because $\overline{X}$ has never been disturbed by any of the
evolutions we performed, it is still a single-body operator localized to the
qubit inside the defect. This leaves the encoded qubit prone to decohering
environmental interactions, and so we make it larger by ``splitting'' the
defect apart into a pair of defects, as depicted in
Fig.~\ref{fig:magicprep6}.
As we move the parts away from each other, we also grow their perimeters
using the methods described above to protect against $\overline{Z}$ errors.

\begin{figure}[!htb]
\centering
\includegraphics[width=0.95\columnwidth]{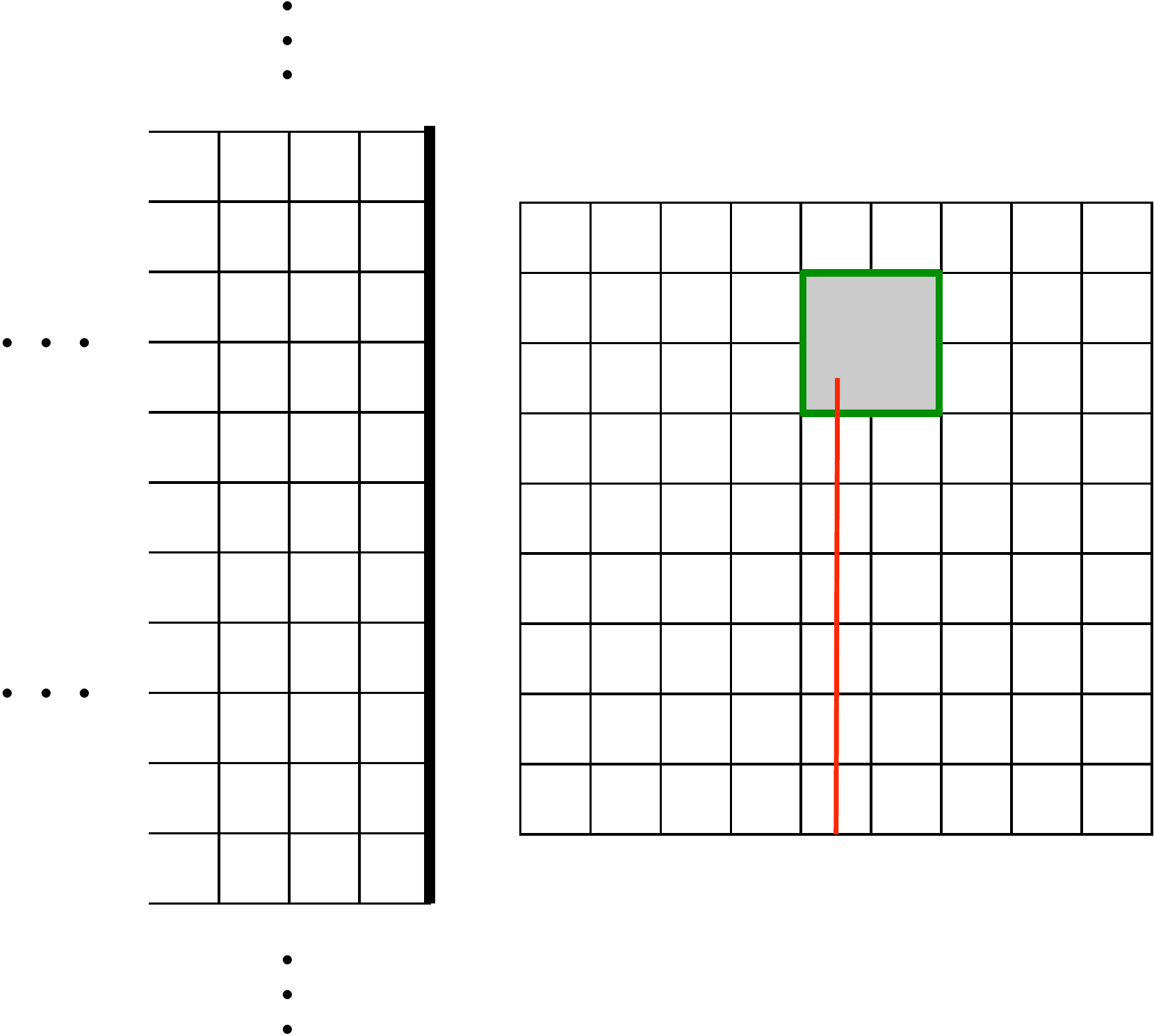}
\caption{One of the defects is merged with the boundary to make the standard
single defect.\label{fig:magicprep6merged}}
\end{figure}

This double-defect qubit could be used as-is, but to make it more like the
defects we have worked with so far, we simply take one of the halves and merge
it with the global smooth boundary of our preparation region, as depicted in
Fig.~\ref{fig:magicprep6merged}.
Finally, we attach the surface containing this defect to the main surface
using the procedure described in Sec.~\ref{sec:isolate}. This defect can be
shuttled in and the boundary can be modified to the original shape.

\begin{figure}[!htb]
\centering
\includegraphics[width=0.95\columnwidth]{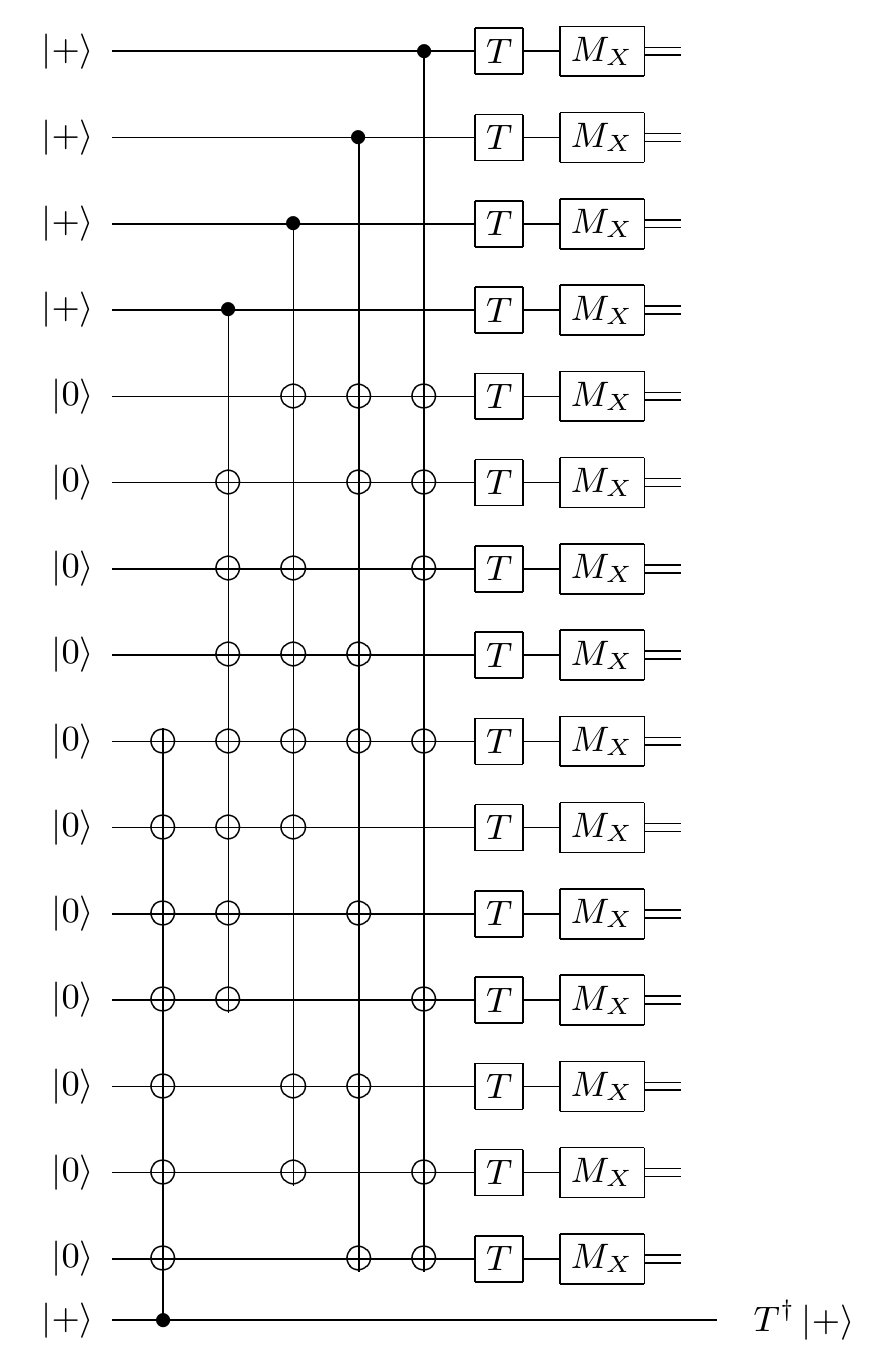}
\caption{Distillation circuit for $T |+\>$ states, constructed from the
15-qubit Reed-Muller code's encoding circuit.\label{fig:bk_distill}}
\end{figure}

\begin{figure}
\centerline{
\Qcircuit @C=1em @R=1em { \lstick{\ket{+}} & \qw       & \ctrl{6} & \qw &
\qw      & \gate{S} & \gate{M_X} & \cw \\ \lstick{\ket{+}} & \qw       & \qw
& \ctrl{5} & \qw      & \gate{S} & \gate{M_X} & \cw \\ \lstick{\ket{+}} &
\qw       & \qw      & \qw      & \ctrl{3} & \gate{S} & \gate{M_X} & \cw \\
\lstick{\ket{0}} & \qw       & \targ    & \targ    & \targ    & \gate{S} &
\gate{M_X} & \cw \\ \lstick{\ket{0}} & \targ     & \qw & \targ    & \targ
& \gate{S} & \gate{M_X} & \cw \\ \lstick{\ket{0}} & \targ     & \targ    &
\qw      & \targ    & \gate{S} & \gate{M_X} & \cw \\ \lstick{\ket{0}} &
\targ     & \targ    & \targ    & \qw      & \gate{S} & \gate{M_X} & \cw \\
\lstick{\ket{+}} & \ctrl{-3} & \qw      & \qw      & \qw & \qw      & \qw
& \qw & \rstick{S^\dagger \ket{+}} } }
\caption{A distillation protocol for $S\ket{+}$ states based on the encoding
circuit for the $\lbrack \! \lbrack 7, 1, 3\rbrack \! \rbrack$ quantum
Steane code.\label{fig:steane_distill}}
\end{figure}
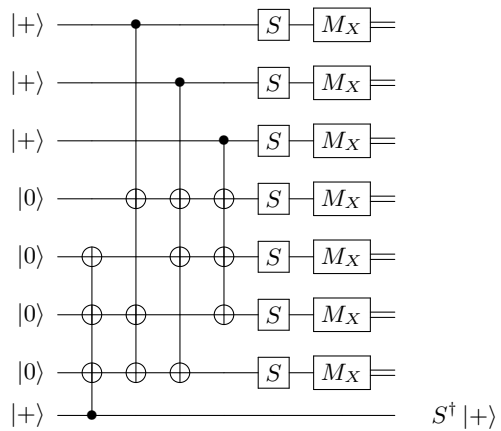

Encoded distillation circuits, such as the ones depicted in
Figs.~\ref{fig:bk_distill} and \ref{fig:steane_distill} (where the $S$ and
$T$ gates are implemented by teleportation circuits such as the one depicted
in Fig.~\ref{fig:gate_teleport}), utilize operations we have already
described: preparation of defects in $|0\>$, $|+\>$, $T|+\>$, and $S|+\>$
states and implementation of $CNOT$ gates by code deformation.  The only
encoded operations these circuits use that we have not described are
measurements of encoded Pauli $\overline{X}$ and $\overline{Z}$ operators,
which we describe in Sec.~\ref{sec:pauli_meas}.

%%%%%%%%%%%%%%%%%%%%%%%%%%%%%%%%%%%%%%%%%%%%%%%%%%%%%%%%%%%%%%%%%%%%%%%%%%%%%%%
% Section
%
\section{Non-adiabatic procedures for surface code defects}
\label{sec:non_adiabatic}

The procedures presented in Sec.~\ref{sec:acd} use only adiabatic evolutions
of stabilizer Hamiltonians. However, these operations do not allow for
universal quantum computation. The key missing ingredient is the capability
to perform logical measurements---namely, the ability to measure
$\overline{X}$ and $\overline{Z}$ for smooth and rough defects. These
measurements are the only non-adiabatic ingredients appearing in our model.
In this section we describe how to perform them as well as use them in
additional procedures, such as heralded application of $\overline{X}$ and
$\overline{Z}$ gates. Although the measurements are not protected by
adiabaticity or a Hamiltonian gap, their topological nature provides
robustness to local errors.

%%%%%%%%%%%%%%%%%%%%%%%%%%%%%%%%%%%%%%%%%%%%%%%%%%%%%%%%%%%%%%%%%%%%%
% Subsection
%
\subsection{Measurements of \texorpdfstring{$\overline{X}$}{logical X} and \texorpdfstring{$\overline{Z}$}{logical Z} for defects}
\label{sec:pauli_meas}

In measurement-based surface-code models, defect logical operators are
measured in-situ by simply measuring a region of individual qubits in the
surface. The parities of of the measurements are then used to infer the
eigenvalue of $\overline{X}$ or $\overline{Z}$ with probability $1 -
O(p^d)$, where $d$ is the distance of the code and $p$ is the probability
that an individual qubit measurement is faulty. In our Hamiltonian model,
this in-situ measurement is an issue because single-qubit measurements in
the surface will necessarily anti-commute with the code Hamiltonian, leading
to excitations out of the ground space. If it is the end of the computation,
and we want to know the state of all the defect qubits, we can just turn the
Hamiltonian off and measure everything. However, the use of magic states via
gate teleportation (described later) requires conditioning future actions on
the classical outcome of logical qubit measurements. In this section we
present an ancilla-coupled method to perform these logical measurements.

\begin{figure}
\centering
$$\Qcircuit @C=1.5em @R=1.5em {
\lstick{\ket{\psi}_{\text{smooth}}} & \ctrl{1} & \qw
\\
\lstick{\ket{0}_{\text{rough}}}     & \targ    & \measureD{Z}
}$$
\caption{An example of measuring $\overline{Z}$ for a smooth qubit. It
requires the preparation of a rough defect in a $+1$ eigenstate of
$\overline{Z}$, as discussed in
Sec.~\ref{sec:esprep}.\label{fig:ancilla_coupled}}
\end{figure}
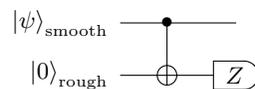

To measure $\overline{X}$ or $\overline{Z}$ for a defect in a
``non-destructive'' way (meaning that the post-measured state stays in the
codespace), we use the method of ancilla-coupled measurement introduced by
Steane in Ref.~\cite{Steane:1996b}. Fig.~\ref{fig:ancilla_coupled} depicts
this process for measuring $\overline{Z}$ for a smooth defect qubit in the
state $|\psi\>$.
First, we prepare a rough defect in the $+1$ eigenstate of $\overline{Z}$ as
described in Sec.~\ref{sec:esprep}. Next, we perform a sequence of adiabatic
deformations, described in Sec.~\ref{sec:deform}, to enact a $CNOT$ gate
between the smooth and rough defects. Then, the rough-defect ancilla is
detached from the code using the method demonstrated in
Sec.~\ref{sec:isolate}. Finally, we turn off the Hamiltonian and
destructively measure the isolated region in the $Z$ basis. A similar
procedure performs a measurement of $\overline{X}$ for a smooth qubit
(simply measure the isolated region in the $X$ basis), and a similar circuit
can be used to measure logical operators for a rough defect.

%%%%%%%%%%%%%%%%%%%%%%%%%%%%%%%%%%%%%%%%%%%%%%%%%%%%%%%%%%%%%%%%%%%%%
% Subsection
%
\subsection{Heralded application of \texorpdfstring{$\overline{X}$}{logical X} and \texorpdfstring{$\overline{Z}$}{logical Z} to
defects}
\label{sec:heralded_pauli}

\begin{figure}[h]
$$\Qcircuit @C=1.5em @R=1.5em {
\lstick{\ket{\psi}_{\text{smooth}}} & \ctrl{1} & \gate{M_{\overline{X}}} &
\ctrl{1} & \rstick{\overline{X}^{a} \overline{Z}^{b} \ket{\psi}} \qw \\
\lstick{\ket{0}_{\text{rough}}}     & \targ    & \qw                     &
\targ    & \gate{M_{\overline{Z}}} 
}$$
\begin{center}
\caption{Circuit used to apply one of the Pauli operators to a smooth defect
qubit. The outcome of the $X$ measurement is $b \in \{0,1\}$ and the outcome
of the $Z$ measurement is $a \in \{0,1\}$. The outcomes of the measurement
all occur with equal probability and the final state depends on these
outcomes as shown.  If an undesired operator is applied, the ancilla qubit
is reinitialized and the circuit is implemented again. However, now the
appropriate operator is the one that undoes the operator applied in the
first iteration {\em and} applies the desired operator. (This, of course,
will just be a different one of the four operators $\overline{X}^a
\overline{Z}^b$.)} 
\label{fig:xgate}
\end{center}
\end{figure}
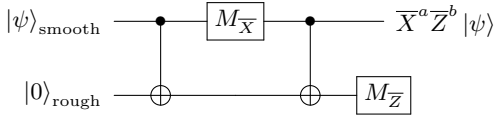

With the ability to perform ancilla-coupled measurements, introduced in
Sec.~\ref{sec:pauli_meas}, and the Hamiltonian evolutions described in
Sec.~\ref{sec:acd}, we can apply $\overline{X}$ and $\overline{Z}$ to
defects using the circuit shown in Fig.~\ref{fig:xgate}, where the
measurements are assumed to be of the type described in the previous
section.  These operations are not necessary to establish universality; the
set of encoded operations we have presented thus far are a universal set by
themselves.  In fact, there is never a need to apply logical Pauli operators
at all using our encoded gate basis because logical Pauli operators can be
propagated through encoded circuits efficiently by the Gottesman-Knill
theorem \cite{Gottesman:1999b}---the only non-Clifford gate in our gate
basis is the preparation $T|+\>$, and Pauli operators never need to be
propagated through preparations.  The propagated ``Pauli frame'' can then be
used to reinterpret measurement results as needed, without active
application of logical Pauli operators.  Nevertheless, we present methods
for applying logical Pauli operators in case there is a situation where
propagating the Pauli frame is undesirable.
 
All of the pieces in this circuit have been described previously. The
preparation of a rough defect in the $+1$ eigenstate of $\overline{Z}$ is
described in Sec.~\ref{sec:esprep}, performing a $CNOT$ between a smooth
defect and a rough defect is described in Sec.~\ref{sec:deform}, and making
measurements of $\overline{X}$ and $\overline{Z}$ for smooth and rough
defects was just described in Sec.~\ref{sec:pauli_meas}.

%%%%%%%%%%%%%%%%%%%%%%%%%%%%%%%%%%%%%%%%%%%%%%%%%%%%%%%%%%%%%%%%%%%%%%%%%%%%%%%
% Section
%
\section{The completed model}
\label{sec:completed_model}

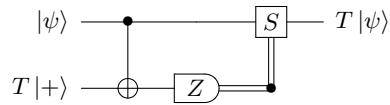
\begin{figure}[h]
$$\Qcircuit @C=1.5em @R=1.5em {
\lstick{\ket{\psi}} & \ctrl{1} & \qw          & \gate{S} &
\rstick{T\ket{\psi}} \qw \\
\lstick{T\ket{+}}   & \targ    & \measureD{Z} & \control \cw \cwx
}$$
\begin{center}
\caption{Gate teleportation circuit using the $T\ket{+}$ state. The $S$
correction needs to be performed half of the time and can be implemented in
the same way using the state $S\ket{+} = \ket{+i}$ instead of $T\ket{+}$
(and utilizing a $Z$ correction half of the time).\label{fig:gate_teleport}}
\end{center}
\end{figure}

\begin{figure}[h]
$$\Qcircuit @C=1.5em @R=1.5em {
\lstick{\ket{\psi}} & \gate{S} & \targ     & \gate{S^\dagger} & \qw
& \gate{A}  & \rstick{H\ket{\psi}} \qw \\
\lstick{\ket{+}}    & \qw      & \ctrl{-1} & \gate{S}         & \measureD{X}
& \control \cw  \cwx
}$$
\begin{center}
\caption{Circuit for applying the Hadamard gate with an ancilla state. The
correction $A$ depends on the result of the measurement: if the measurement
result is $+1$, then $A = X$, and if the measurement result is $-1$, then $A
= Z$. The $S$ and $S^\dagger$ gates can be performed using a circuit like
the one in Fig.~\ref{fig:gate_teleport}.\label{fig:hadamard_ancilla}}
\end{center}
\end{figure}
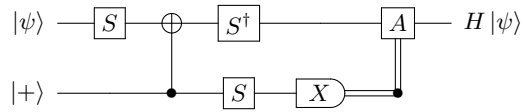

To summarize our surface code model, we list the procedures we have defined
in Sec.~\ref{sec:acd} and Sec.~\ref{sec:non_adiabatic}:
\begin{enumerate}
\item{Sec.~\ref{sec:surface_creation}: Adiabatic preparation of a surface
code encoding no qubits}
\item{Sec.~\ref{sec:create}: Adiabatic preparation of smooth defects in the
$+1$ eigenstate of $\overline{Z}$ and rough defects in the $+1$ eigenstate
of $\overline{X}$}
\item{Sec.~\ref{sec:deform}: Adiabatic deformation of smooth and rough
defects, allowing for defect movement} 
\item{Sec.~\ref{sec:isolate}: Adiabatic detaching and attaching procedures,
allowing for the isolation of regions containing defects}
\item{Sec.~\ref{sec:esprep}: Adiabatic preparation of smooth defects in the
$\pm 1$ eigenstate of $\overline{X}$ and rough defects in the $\pm 1$
eigenstate of $\overline{Z}$}
\item{Sec.~\ref{sec:magic}: Adiabatic injection of ancilla states into
defects}
\item{Sec.~\ref{sec:pauli_meas}: Non-adiabatic procedures for
``non-destructive'' ancilla-coupled measurement of $\overline{X}$ and
$\overline{Z}$ for defects}
\item{Sec.~\ref{sec:heralded_pauli}: Non-adiabatic, measurement-based
procedure for the heralded application of $\overline{X}$ and $\overline{Z}$}
\end{enumerate}
Magic-state gate teleportation of the $T$ gate is performed using the
circuit in Fig.~\ref{fig:gate_teleport}, and the Hadamard gate can be
performed with an ancilla state using the circuit in
Fig.~\ref{fig:hadamard_ancilla}.
In both cases, the only operations required involve the procedures defined
in the list above. Other procedures, such as performing a $CNOT$ between two
smooth qubits, have been studied previously \cite{Raussendorf:2007b} and
also only require operations from the list above. Thus, in encoded form, we
can prepare Pauli $\overline{X}$ and $\overline{Z}$ eigenstates, perform a
universal gate set, and measure any qubit in either the $\overline{X}$ or
$\overline{Z}$ basis.  Taken together, these procedures allow for universal
quantum computation.

%%%%%%%%%%%%%%%%%%%%%%%%%%%%%%%%%%%%%%%%%%%%%%%%%%%%%%%%%%%%%%%%%%%%%%%%%%%%%%%
% Section
%
\section{Extension to 2D color codes}
\label{sec:color}

We briefly discuss how one can adapt our surface code procedures to the
two-dimensional color codes, in particular to the $4.8.8$ $2$D color code.
This extends our construction to all nontrivial homological stabilizer
codes, because by Anderson's classification theorem \cite{Anderson:2011a},
all homological stabilizer codes with nonlocal logical operators are either
surface codes or color codes.

\begin{figure}[h]
\begin{center}
\includegraphics[width=0.95\columnwidth]{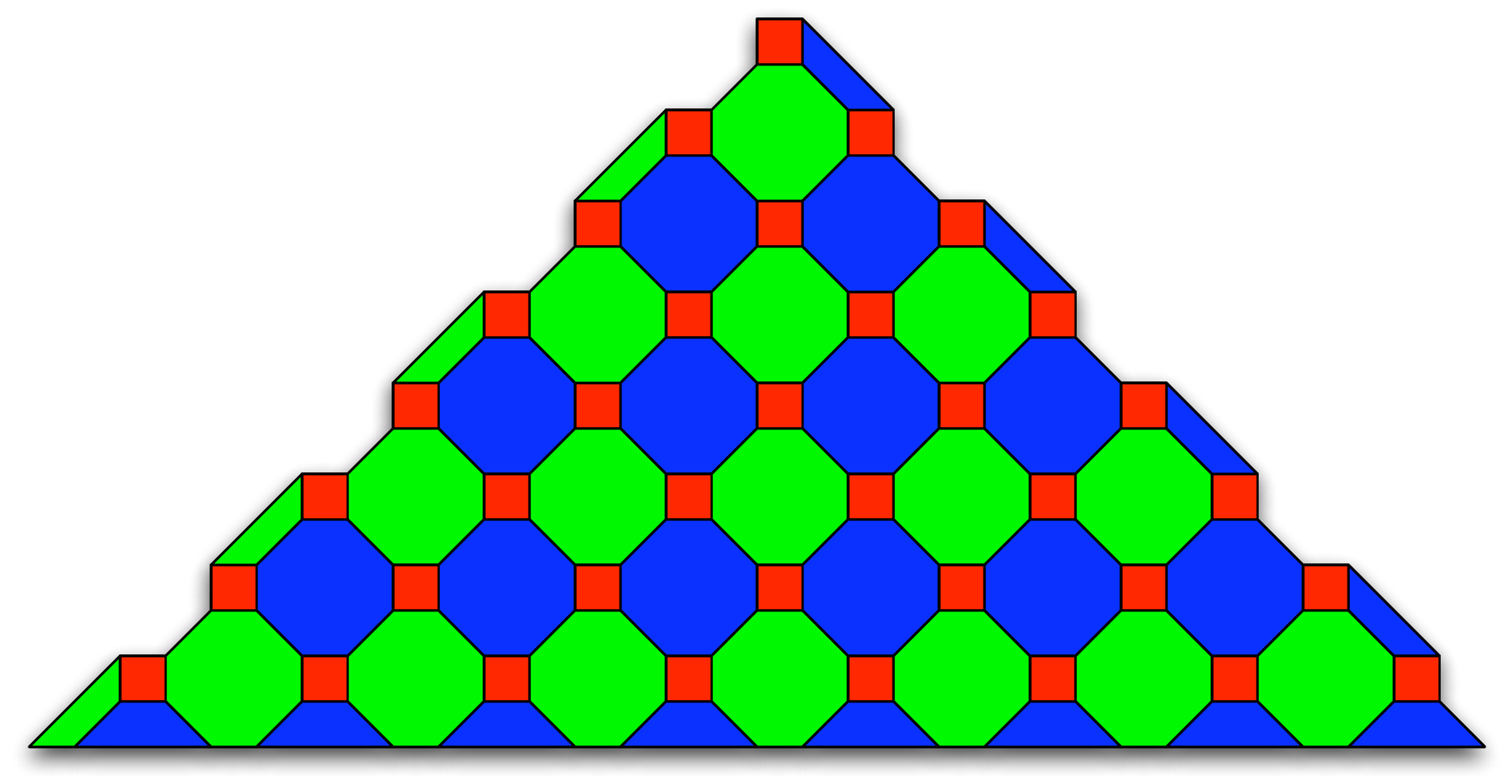}
\caption{A lattice with colored plaquettes on which one can define the color
codes. }
\label{fig:colorlattice}
\end{center}
\end{figure}

Color codes in two dimensions are defined on a two-dimensional lattice that
is trivalent (each vertex is of degree three) and face-three-colorable (we
can color the plaquettes by three colors such that no two adjacent
plaquettes are the same color).  In such a graph, the edges can also be
colored to be the color that is different from the colors of the two faces
incident upon it.  Fig.~\ref{fig:colorlattice} is an example of
such a lattice.
Unlike our presentation of the surface code in which graph edges were
associated with qubits, color codes are naturally presented so that graph
vertices are associated with qubits.  Let $V(p)$ denote the vertices that
are on the boundary of a plaquette, and define a stabilizer group structure
of the color codes as follows. To every plaquette $p$, associate two
stabilizer generators, the tensor product of Pauli $X$ on the adjacent
qubits, given by
\begin{equation}
S_p^{X}=\bigotimes_{v \in V(p)}X_v,
\end{equation}
as well as the tensor product of Pauli $Z$ on the adjacent qubits, given by  
\begin{equation}
S_p^{X}=\bigotimes_{v \in V(p)} Z_v.
\end{equation}
The representative code in Fig.~\ref{fig:colorlattice} has four-body (red)
and eight-body (blue and green) stabilizer generators. (These are the
weights away from the boundaries of the code, where four-body blue and green
faces also exist.) Boundaries in the color code also have a slightly richer
structure. They are no longer smooth and rough, but rather, they have a
color associated to them. This color is determined by the boundary's {\em
missing color}. For example, in Fig~\ref{fig:colorlattice}, the bottom
boundary is red, since there are no red plaquettes adjacent to the bottom
edge. A careful accounting of qubits and checks in
Fig.~\ref{fig:colorlattice} indicates that there is a single logical qubit
associated with the surface. For our purposes, we will treat it as a
``gauge'' degree of freedom using the subsystem stabilizer code formalism
\cite{Bacon:2006a}. The operators $\overline{X}$ and $\overline{Z}$
associated with this qubit can be chosen as strings of Pauli $X$ and $Z$
operators, respectively, along the bottom boundary.

Just as with the surface codes, we can create defects in the color code to
store more logical qubits. In addition to having a type ($X$ or $Z$), the
defects now also have a color. To create the analog of a smooth defect, we
remove a $Z$-type stabilizer generator, and to create the analog of a rough
defect, we remove an $X$-type generator. For a $Z$-type defect, one choice
for $\overline{Z}$ is the removed generator (equivalent to a string of a
{\em different color} around the defect that only passes through edges and
faces of that color), and one choice for $\overline{X}$ is a string of $X$s
connecting to a boundary whose color is the same as that of the removed
plaquette (such that the string only passes through edges and faces of the
{\em same color} as the removed plaquette).

As is true for any stabilizer code, we can define the Hamiltonian in
Eq.~(\ref{eq:ham}), and it has a ground space equivalent to the codespace of
the code. In the case of the color codes it can be written as
\begin{equation}
\label{eq:color_ham}
H = - \sum_p \left( S_p^X + S_p^Z \right).
\end{equation}
The color-code Hamiltonian, like the surface-code Hamiltonian, does not lead
to a self-correcting quantum memory, but we can use adiabatic interpolations
between static Hamiltonians of the type in Eq.~(\ref{eq:color_ham}).

\begin{figure}[!htb]
\centering
\includegraphics[width=0.95\columnwidth]{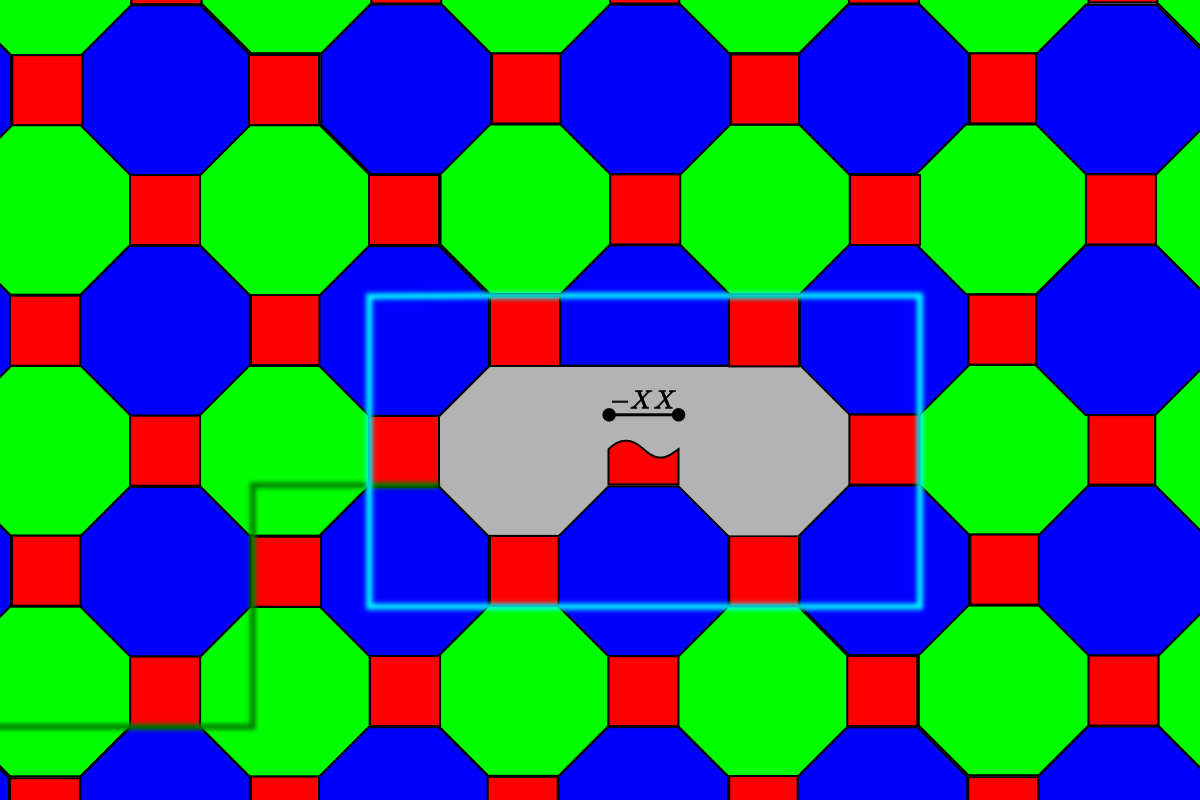}
\caption{Two adjacent green $Z$-plaquettes are turned off while turning on
the $-XX$ Hamiltonian shown, creating a $Z$-type defect in the $+1$
eigenstate of $\overline{Z}$ (shown here as a light blue string encircling
the defect).  The green string depicts the associated $\overline{X}$
operator.\label{fig:color_defect}}
\end{figure}

As in Sec.~\ref{sec:surface_creation}, we can perform an adiabatic
interpolation to initially create the color code without any defects. We
imagine the same setting---a large number of qubits in the ground state of
local Hamiltonians $H = -Z$---and prepare the code by using an interpolation
of the form
\begin{align}
\nonumber
H(t)
 &= \left(1-\frac{t}{T}\right) \sum_{j \in \mathcal{Q}}
        \left( -Z_j\right)
  + \frac{t}{T} \sum_p \left(- S_p^X - S_p^Z \right) \\ 
  &+ \frac{t}{T} \sum_{j \not \in \mathcal{Q}} \left( - Z_j \right).
\end{align}
(Since $\overline{Z}$ for the newly created code commutes with this
Hamiltonian at all times, and since it initially has eigenvalue $+1$, the
qubit associated with the surface is prepared in the $+1$ eigenstate of
$\overline{Z}$. This is the gauge degree of freedom mentioned above.) As we
did for the surface code, we choose to create a small color code first and
then grow it to avoid a shrinking gap. The small color code is grown in a
manner similar to Sec.~\ref{sec:surface_creation}.  For example, to create a
green $Z$-type defect in the $+1$ eigenstate of $\overline{Z}$, described
for the surface code in Sec.~\ref{sec:create}, two $Z$-type green plaquettes
separated by one red plaquette are turned off while simultaneously turning
on $-XX$ on two pairs of qubits in between, as shown in
Fig.~\ref{fig:color_defect}.
Note that during the defect's creation a neighboring blue plaquette gets
modified to a six-body operator and a neighboring red plaquette gets
modified to a two-body operator.
% Additionally, the $X$-type plaquette
%operators located at the same place as the defect 

The surface code procedures for growing and moving defects, presented in
Sec.~\ref{sec:deform}, can also be adapted to the color codes. We will not
present the the cases for different numbers of interior qubits separately
here. Rather, we examine the simplest case when there are only two
neighboring qubits. The other cases, as in the surface code, simply require
more modifications of adjoining checks. To grow a $Z$-type green defect like
the one in Fig.~\ref{fig:color_defect}, first pick another green face. It
will be separated from the defect region by a red plaquette. Along one of
the two lines connecting the defect region to the green check, turn on $-XX$
while turning off the green plaquette. This will incur a modification a
neighboring blue plaquette as well as the red plaquette itself.

\begin{figure}[!htb]
\centering
\includegraphics[width=0.95\columnwidth]{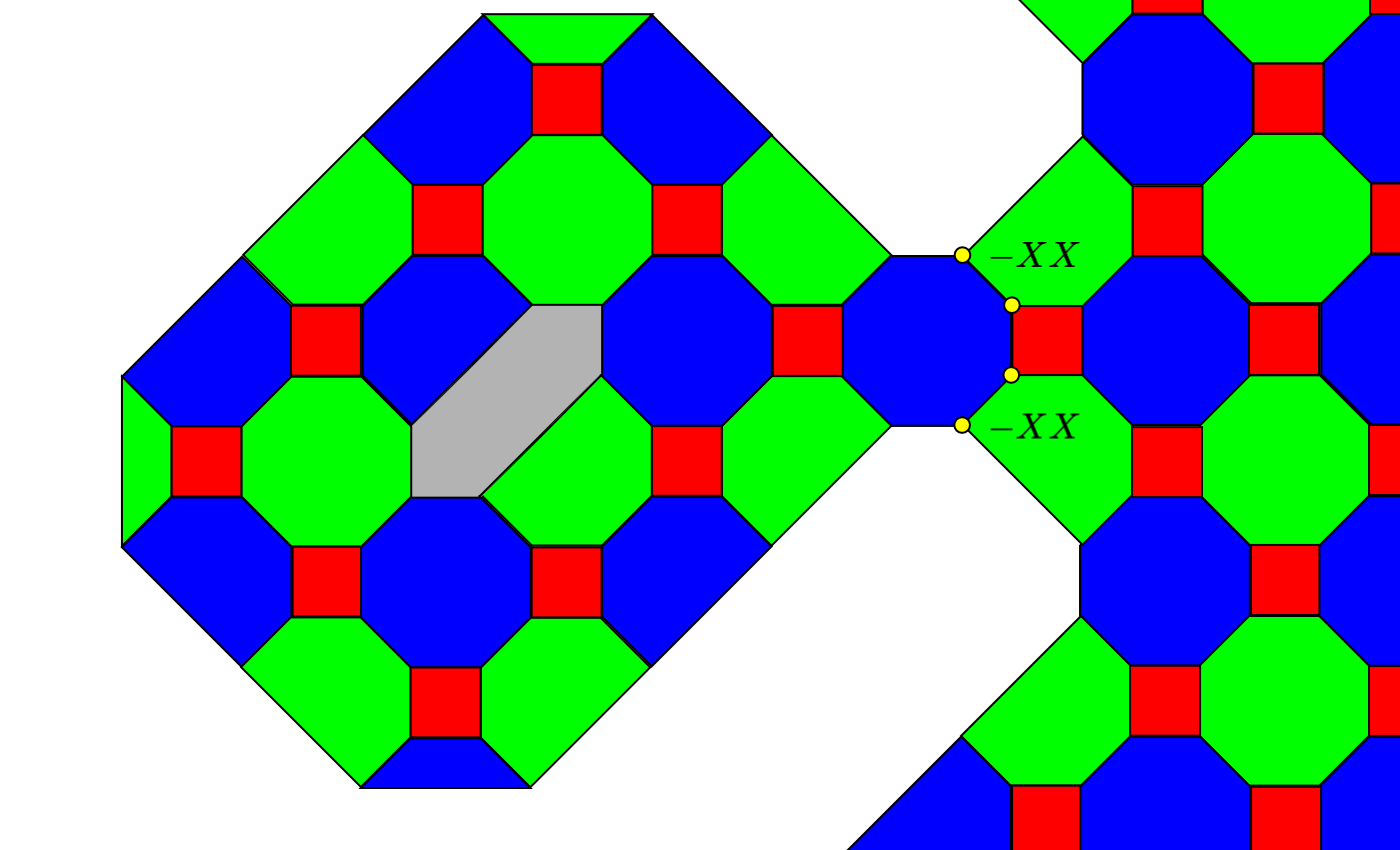}
\caption{A $Z$-type red defect isolation procedure. The ``drawbridge'' in
this case is the red plaquette adjacent to the yellow dots in the figure.
The $Z$-type check on the red face is turned off while the two $-XX$
operators are turned on. The four-body $X$ operator that is the product of
the two $-XX$ Hamiltonians is in the stabilizer group before the evolution,
and it is trivially in the stabilizer group of the code after the evolution.
The blue and green plaquettes adjacent to the yellow dots are modified to be
a four-body operators. (Also note that the $X$-type check on the red
plaquette must also turned off to fully isolate the region, and two $-ZZ$
Hamiltonians are turned on.)\label{fig:color_isolate}}
\end{figure}

Next, we show that the color code also supports detachment and attachment
procedures, described in Sec.~\ref{sec:isolate} for the surface code.
Imagine a two-plaquette red defect, depicted in
Fig.~\ref{fig:color_isolate}, that we would like to isolate from the bulk
code.
To complete the detachment procedure for a $Z$-type red defect, two $-XX$
Hamiltonians---on the qubits indicated by yellow dots---are turned on while
turning off the $Z$-type red plaquette operator adjacent to the dots. In the
process, the adjacent blue and green plaquettes get modified to four-body
operators. Since the four-body $X$ operator that is the product of the two
$-XX$ Hamiltonians is in the stabilizer group at the beginning {\em and} at
the end of the evolution, we have successfully severed the two code regions.

As discussed in Sec.~\ref{sec:esprep}, it is important that we are able to
prepare $Z$-type defects in eigenstates of $\overline{X}$ and vice versa.
For color codes, the procedure is essentially identical to the one for
surface codes, and proceeds by preparing single qubits in particular states
($\pm 1$ eigenstates of $X$ for $Z$-type defects and $\pm 1$ eigenstates of
$Z$ for $X$-type defects). Just as before, a defect location is anticipated
and the preparation of the surface proceeds normally everywhere except for
the defect.

\begin{figure}[!htb]
\centering
\includegraphics[width=0.95\columnwidth]{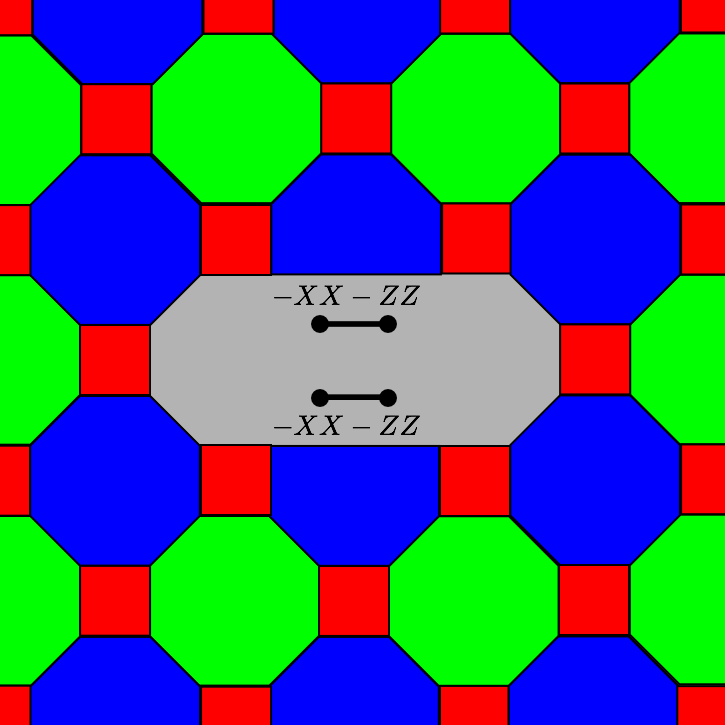}
\caption{The creation of a defect region with both the $X$-type and $Z$-type
green checks turned off. There are four interior qubits prepared in two Bell
pairs by this procedure.\label{fig:488inject2}}
\end{figure}

\begin{figure}[!htb]
\centering
\includegraphics[width=0.95\columnwidth]{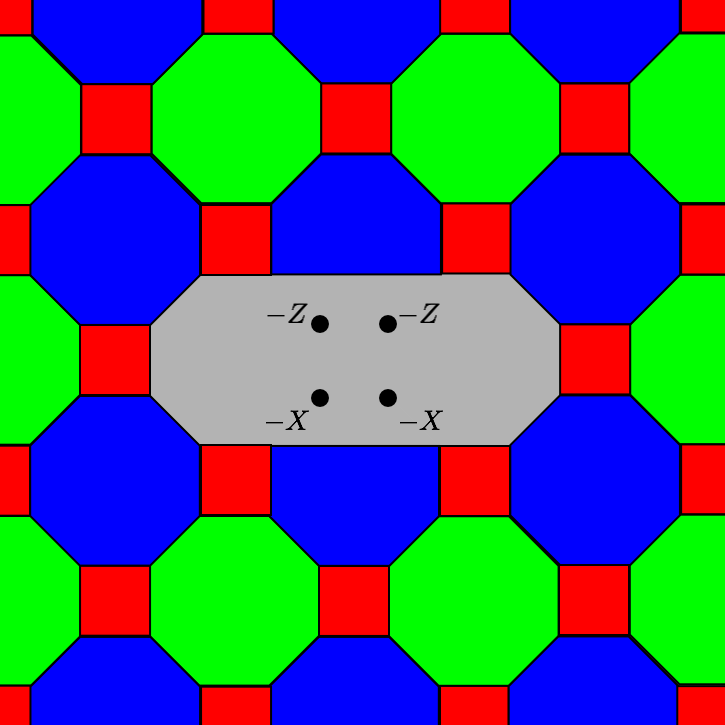}
\caption{Four interior qubits are ``exposed.''\label{fig:488inject3}}
\end{figure}

\begin{figure}[!htb]
\centering
\includegraphics[width=0.95\columnwidth]{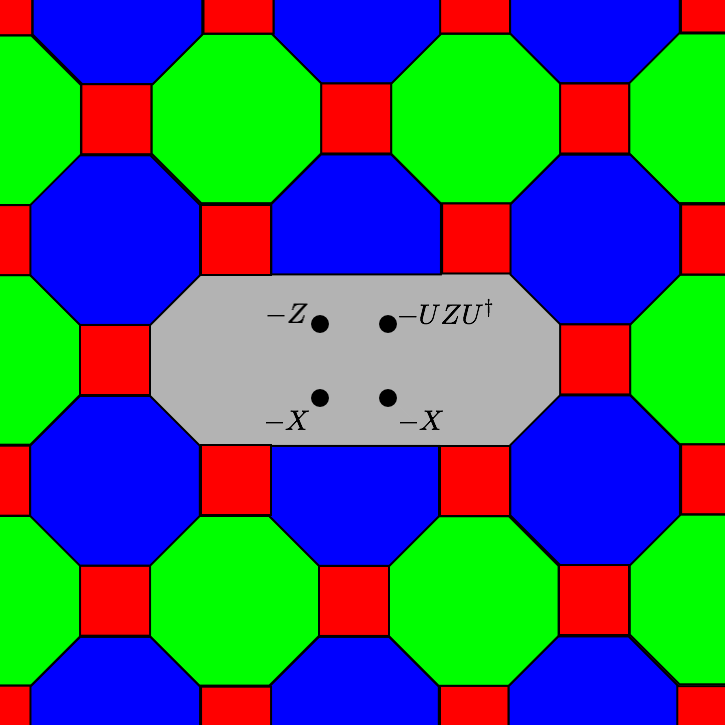}
\caption{The upper-right qubit is adiabatically dragged to the desired
state. For instance, to inject $T\ket{+}$ states,
$U=TH$.\label{fig:488inject4}}
\end{figure}

\begin{figure}[!htb]
\centering
\includegraphics[width=0.95\columnwidth]{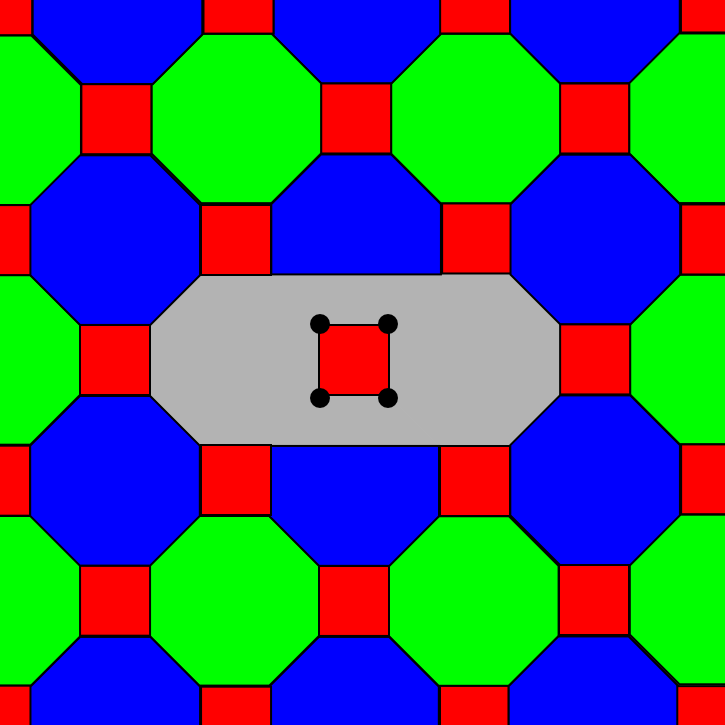}
\caption{The single-body terms in Fig.~\ref{fig:488inject4} are turned off
while turning on the $X$-type and $Z$-type checks on the red
plaquette.\label{fig:488inject5}}
\end{figure}

\begin{figure}[!htb]
\centering
\includegraphics[width=0.95\columnwidth]{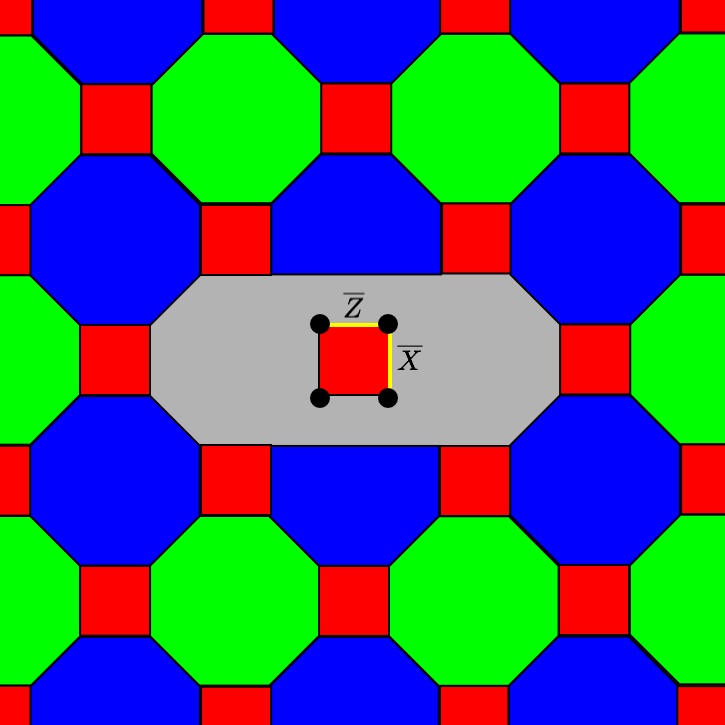}
\caption{$\overline{X}$ and $\overline{Z}$ after the reintroduction of the
red plaquette in Fig.~\ref{fig:488inject5}.\label{fig:488inject6}}
\end{figure}

\begin{figure}[!b]
\centering
\includegraphics[width=0.95\columnwidth]{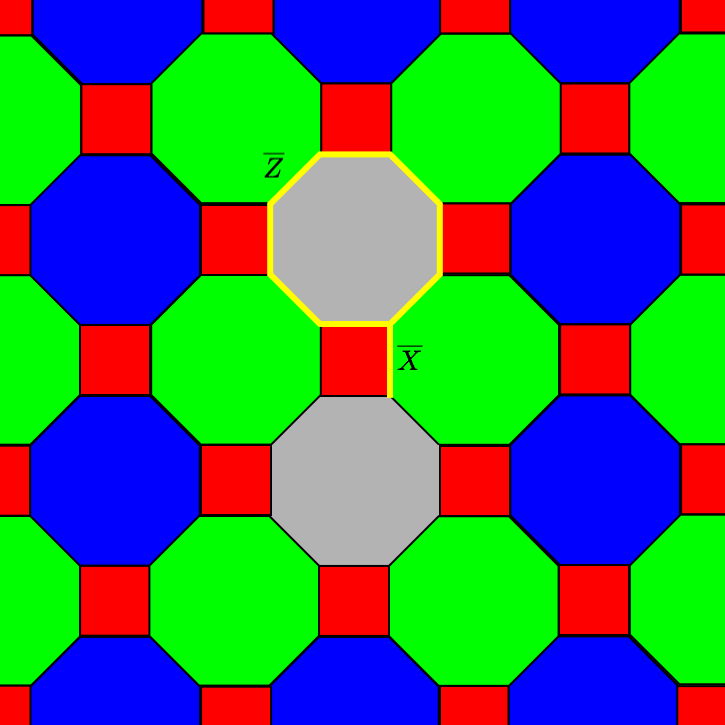}
\caption{The arrangement of the defect after reintroducing the $X$-type
green plaquettes. $\overline{X}$ is a string of Pauli $X$ operators
connecting two blue faces and $\overline{Z}$ is a loop of Pauli $Z$
operators around a blue face.\label{fig:488inject7}}
\end{figure}

Ancilla state injection for the color codes is slightly different than the
procedures for the surface code introduced in Sec.~\ref{sec:magic}.  After
isolating a region with green boundaries, or creating such a region adjacent
to a green boundary, we use the procedures described above to introduce an
$X$-type {\em and} a $Z$-type defect at the same location, as pictured in
Fig.~\ref{fig:488inject2}.
Notice that the interior red checks have also been modified during this
procedure, putting the four interior qubits into two Bell pairs.
Additionally, the neighboring blue plaquettes have been modified to six-body
operators. An evolution is then performed that only touches these four
interior qubits, turning on the Hamiltonians pictured in
Fig.~\ref{fig:488inject3} while turning off the two $-XX-ZZ$ Hamiltonians.
Next, just as we did for the surface code, we adiabatically drag a qubit to
the desired state, as pictured in Fig.~\ref{fig:488inject4}.
The ``logical qubit'' is localized to the upper-right qubit, with
single-body $\overline{X}$ and $\overline{Z}$ operators. The next step is to
``grow'' these logical operators in a particular way. This is achieved by
performing another adiabatic evolution on the four qubits to the Hamiltonian
represented in Fig.~\ref{fig:488inject5}, which is just the reintroduction
of the red face checks that we turned off at the beginning.
This evolution modifies $\overline{X}$ and $\overline{Z}$ from single-body
operators to the operators shown in Fig.~\ref{fig:488inject6}.
Finally, the $X$-type checks on the green faces currently housing the defect
are turned on while the adjacent $Z$-type blue faces are turned off, leading
to the situation depicted in Fig.~\ref{fig:488inject7}.
As in the case of the surface code, one of these faces is moved away and
absorbed into the green boundary of the region. Then the region is attached
and the green defect encoding the state is moved into the bulk computational
region.

None of the other procedures introduced in Sec.~\ref{sec:acd} and
Sec.~\ref{sec:non_adiabatic} are appreciably different for the color codes.
Measurements are still performed in an ancilla-coupled manner, and
$\overline{X}$ and $\overline{Z}$ can still be applied in a heralded
fashion. Logical $CNOT$ gates are still performed by braiding, with the
control being a $Z$-type defect and the target being an $X$-type defect.
Ref.~\cite{Landahl:2011a} discusses how to perform a $CNOT$ between defects
of the same type (or color). Thus, all the ingredients are precisely the
same, and encoded universal quantum computation can be performed with two
ingredients: adiabatic interpolations between static Hamiltonians and
ancilla-coupled measurements.

%%%%%%%%%%%%%%%%%%%%%%%%%%%%%%%%%%%%%%%%%%%%%%%%%%%%%%%%%%%%%%%%%%%%%%%%%%%%%%%
% Section
%
\section{Conclusion}

We have presented a model of quantum computation that utilizes adiabatic
interpolations between static Hamiltonians which encode quantum information
in their degenerate ground spaces. By utilizing the process of adiabatic
code deformation, we create and grow small code regions, introduce and braid
defects, and inject arbitrary states into defects. These
procedures never cause the Hamiltonian gap to shrink below a constant
proportional to $\Delta$, and they can all be performed with the protection
of a gap and topology. However, to perform logical measurements we use an
ancilla-coupled scheme, braiding and isolating an ancilla defect and then
turning pieces of the Hamiltonian off and destructively measuring a code
region. Taken together, these procedures allow for universal quantum
computation.

Our model lives at the intersection of three other models of quantum
computation. It provides explicit examples of adiabatic evolutions in the
setting of a topological code, and we make an effort to supply procedures
that do not increase the rate at which errors (anyons) are introduced to the
system. Since we store information in the ground space of a changing
Hamiltonian, our model also borrows intuition and robustness from holonomic
quantum computing. Indeed, the braiding operations we perform rely precisely
on the non-trivial structure of ground space holonomies. Lastly, our
adiabatic interpolations are like miniature adiabatic quantum computations,
and their implementations are made less noisy by traversing an adiabatic
path more slowly.

Unfortunately, the model we present is not fault-tolerant. While the
lifetime of the ground space, and thus the encoded quantum information, is
exponential in $\Delta/T$ in the presence of coupling to a thermal bath, no
protection is gained by increasing the size of the code. It would be
interesting to study a model that can actively remove entropy from the
system, utilizing active error correction in a way that is compatible with
the Hamiltonian nature of the model, but we do not address these problems in
this work.

We hope that the model we have analyzed here can be useful for a further
understanding of the properties of quantum computation based on stabilizer
Hamiltonians. In particular, it would be interesting to extend this work to
models such as Kitaev's quantum double model~\cite{Kitaev:1997a} or the
Turaev-Viro codes~\cite{Koenig:2010b}, where universality can be achieved
without the creation and distillation of magic states.

Another line of inquiry worth investigating is the degree to which the control
requirements on our construction can be relaxed.  In particular one can imagine
moving a defect not by turning off and on a few terms in a Hamiltonian to perform a 
deformation, but instead by turning large numbers of these terms off and on at the
same time.  This would have the advantage of not requiring precise few-term
control of a Hamiltonian, but a spatially more course-grained ability to change the 
Hamiltonian.  In \cite{Bacon:2013a} such an approach was investigated for 
adiabatic implementations of measurement-based quantum computing, where
it was argued that this results in a non-constant, but inverse-polynomial energy 
gap.  Such a gap would require a slower evolution to maintain the adiabatic
condition, and one might worry that it would also destroy the robustness of the
mode.  However \cite{Bacon:2013a} argued that this polynomial gap did not 
destroy the error protection properties in a worse manner than the constant gap
model.  Can the topological adiabatic evolutions we describe here be done similarly,
with the ability to only change the Hamiltonian over a spatial course graining?
\\

\noindent \textit{Addendum}: In the course of writing this manuscript, Zheng
and Brun in Ref.~\cite{Zheng:2014a} published an article on a similar topic;
it is worth comparing and contrasting our work to theirs.

Both works bring together concepts from holonomic, adiabatic, and
circuit-model quantum computing to effect universal quantum computation.
Both works also utilize adiabatic interpolations between degenerate
Hamiltonians in a way that maintains a constant energy gap.

Our work differs in that we expand in detail about how adding the extra
ingredient of topological codes into the mix offers additional modes of
error suppression and local quantum processing.  We develop explicit methods
for how adiabatic, holonomic, and circuit-model ideas can be brought to bear
on topological codes, leading to a comprehensive model of ``adiabatic
topological quantum computing.''  The work by Zheng and Brun concludes with
the sentences ``We hope to apply our method to fault-tolerant schemes based
on large block codes and topological codes, which may have higher thresholds
than fault-tolerant schemes that concatenate small codes.  Very likely, the
maximum weight of the Hamiltonian terms used to describe topological codes
during adiabatic evolution will be small and well bounded.''

Our work also differs in that we focus on a model of quantum computation
that only uses (a) adiabatic interpolations between Hamiltonians that can
never be completely turned off and (b) measurements of logical operators; we
further we make it clear that these operations alone are insufficient to
make our model fault-tolerant relative to standard definitions of
fault-tolerance.  The work by Zheng and Brun focuses on a model that has
these operations but also adds (c) the ability to measure code check
operators and (d) the ability to completely turn on and off Hamiltonians.
With these additional features, their model becomes fault tolerant according
to a definition of fault-tolerance they provide.  Augmented with these
capabilities, our model also becomes fault-tolerant according to their
definition, but we have not highlighted this property in the main text as
our emphasis is on the features of pure adiabatic topological quantum
computing model.

Finally, by explicitly going through the steps of how to implement each
element of a universal set of encoded operations, we have found that
contrary to the statement in the work of Zheng and Brun that ``standard
techniques, like magic state injection and distillation, can realize
fault-tolerant encoded non-Clifford gates,'' it can in fact be quite subtle
as to how to realize state injection by adiabatic interpolations.  Indeed,
we point out that there are even qualitative differences between how to
do this correctly for surface codes and for color codes.  Repeating the end of
our main Conclusion section above, this revelation suggests that an
interesting area for future research would be extending our analysis to
models that can achieve universality without the need for the creation and
distillation of magic states.

%%%%%%%%%%%%%%%%%%%%%%%%%%%%%%%%%%%%%%%%%%%%%%%%%%%%%%%%%%%%%%%%%%%%%%%%%%%%%%%
% Acknowledgments (RevTeX Only)
%
\begin{acknowledgments}

CC and AJL were supported in part by NSF grant 0829944.
CC and AJL were supported in part by the Laboratory Directed Research and
Development program at Sandia National Laboratories.  Sandia National
Laboratories is a multi-program laboratory managed and operated by Sandia
Corporation, a wholly owned subsidiary of Lockheed Martin Corporation, for
the U.S.\ Department of Energy's  National Nuclear Security Administration
under contract DE-AC04-94AL85000.
DB and AN were supported in part by NSF grants 0621621, 0803478, 0829937,
and 091640.  DB was supported in part by the DARPA/MTO QuEST program through
grant FA9550-09-1-0044 from AFOSR.
STF was supported by the IARPA MQCO program, by the US Army Research Office
grant numbers W911NF-14-1-0098 and W911NF-14-1-0103, by the ARC via EQuS
project number CE11001013, and by an ARC Future Fellowship FT130101744.

\end{acknowledgments}

\bibliographystyle{landahl}
\bibliography{adiatoric}

\end{document}